\documentclass[12pt]{article}
\usepackage{psfig}

\begin{document}

\begin{titlepage}
\noindent
\begin{flushright}
CERN-TH/2000-385\\
\today
\end{flushright}

\vspace{1cm}
\begin{center}
  \begin{Large}
  \begin{bf}
    Neutrino Masses and Lepton-Flavor Violation in 
    Supersymmetric Models \\with lopsided Froggatt--Nielsen charges
   \\
  \end{bf}
 \end{Large}
\end{center}
  \vspace{1cm}
    \begin{center}
Joe Sato$^a$ and Kazuhiro Tobe$^b$\\
\vspace{0.3cm}
\begin{it}
(a) Research Center for Higher Education, Kyushu University, \\
Ropponmatsu, Chuo-ku, Fukuoka, 810-8560, Japan
\end{it}
\\
      \vspace{0.3cm}
\begin{it}
(b) CERN, Theory Division, CH-1211 Geneva 23, Switzerland \\
\end{it}
  \end{center}
  \vspace{1cm}
\abstract{
We analyze in detail lepton-flavor violation (LFV) in the charged-lepton 
sector such as $\mu \rightarrow e \gamma$, $\tau \rightarrow \mu \gamma$,
$\mu \rightarrow eee$ and the $\mu \rightarrow e$ conversion in nuclei,
within the framework of supersymmetric models with lopsided 
Froggatt--Nielsen charges, in which the large mixing in the neutrino sector 
as well as small mixings in the quark sector can be naturally accommodated.
We show that the present experimental limits on the LFV processes
already exclude some of the models. The future proposed search
for LFV, especially in muon processes, can provide a significant
probe to this framework.

We also stress the importance of the measurement of $U^{MNS}_{e3}$ in neutrino
experiments, and the fact that the KamLAND experiment could play a 
significant role
to test a certain class of models.
}
\end{titlepage}
\section{Introduction}

One of big mysteries in the Standard Model (SM) of elementary particles
is a problem of fermion masses.
Since Yukawa couplings, which
determine the magnitude of fermion masses, are totally free parameters
in the SM,
theoretically we do not know how we can predict a wide variety
of masses of fermions.

On the experimental side,  evidence of non-zero neutrino masses from the
atmospheric neutrino experiment has recently been announced by
the Super-Kamiokande collaboration~\cite{SuperK}.
This result is very interesting because not only it suggests non-zero
neutrino masses, but also it indicates a large mixing in the neutrino 
sector. The tiny but non-zero neutrino masses clearly imply new physics beyond
the SM, and the large mixing in the neutrino sector suggests that a flavor 
structure in the lepton sector seems to be very different from that in 
the quark sector. Therefore, finding the unified picture between small 
mixings in the quark sector and large mixing in the lepton sector 
will be an important key to understanding the problem of 
fermion masses, and a lot of attempts
have been made~\cite{lopsided, neutrino}.

One of the interesting and simple mechanisms to realize the mass
hierarchy of fermions is the Froggatt--Nielsen (FN) mechanism~\cite{FN},
which uses a broken U(1) family symmetry. It has been proposed that
lopsided FN U(1) charges for lepton doublets
would be an interesting candidate to naturally account for the large 
mixing for neutrinos and the small mixings for quarks~\cite{lopsided}.
The low-energy consequence in neutrino physics has been studied
in Ref.~\cite{JY}. It has been also interestingly 
pointed out that this framework can
explain a baryon asymmetry in the present universe~\cite{asaka}.

The supersymmetric (SUSY) extension of this framework,
assuming SUSY is broken at a high-energy scale, is rather
interesting, because we can expect the other low-energy consequence.
The lopsided structure of lepton doublets induces a large mixing in
the Yukawa matrices of leptons. Such a large mixing 
has a potential to
generate the large lepton-flavor violation (LFV) in slepton masses
through the renormalization group (RG) effects~\cite{BM}.
Then LFV processes in the charged-lepton sector such as 
$\mu\rightarrow e \gamma$,
$\tau\rightarrow \mu \gamma$, $\mu \rightarrow eee$, and 
the $\mu \rightarrow e$ conversion in nuclei, are induced 
through diagrams mediated by the sleptons.
In the presence of the large neutrino Yukawa coupling,
the event rates can be within the reach of future experiments
or the models can be strongly constrained~\cite{BM,HMTYY,LFV,JTY}.

In this paper, we analyze the LFV in SUSY models with lopsided FN U(1)
charges in detail.
We show that the search for LFV, especially in muon processes, 
provides a great impact on this framework,
and even at present many of the models are almost excluded.
The future proposed experimental improvement of the search for LFV 
in muon processes 
will be significant for the SUSY models with lopsided
family structure. Therefore, we emphasize
that the LFV search would be an important window to seek for the answer 
to the fermion mass problem. 
In section 2, we briefly introduce the models with lopsided FN U(1)
charges. In section 3, we discuss neutrino masses and mixings in two
interesting classes of models, and especially stress the 
importance of the measurement of the neutrino mixing $U^{MNS}_{e3}$
and the fact that the KamLAND experiment could play a significant role
to test a certain class of models. 
In section 4, we discuss LFV in detail and show that the $\mu \rightarrow
e \gamma$ process is more sensitive to the models than $\tau \rightarrow
\mu \gamma$. Present and future experimental searches have a
great potential to probe the models.

\section{Models with lopsided Froggatt-Nielsen U(1)}
It has been pointed out that lopsided FN U(1) charges for left-handed lepton 
doublets $L_i$~($i=1$--3) are interesting possibilities to 
explain the large $\nu_\mu$--$\nu_\tau$ mixing observed by the atmospheric
neutrino experiments as well as the mass hierarchy of charged leptons and 
quarks. Here we briefly introduce two interesting classes of models.

In order to account for the tiny neutrino masses, we consider the seesaw
mechanism introducing heavy right-handed neutrinos 
$\bar{N}_i~(i=1$--3)~\cite{seesaw}.
One possible interesting U(1) charge assignment is that
the lepton doublets of the second and third families ($L_2$ and $L_3$)
have the same U(1) charges $\tau$ and the first family $L_1$ has a
different U(1) 
charge $\tau+1$, while the right-handed charged leptons $\bar{E}_i$~($i=1$--3)
have U(1) charges 2,~1,~0, respectively. We refer to this class of models
as ``Model I''.
Another interesting charge assignment is that all lepton doublets 
$L_i$ have the same charge $\tau$ and the right-handed charged 
leptons $\bar{E}_i$
have charges 3,~1,~0 respectively. Below, we refer to this class of models
as ``Model II''. We list the FN U(1) 
charges of Models I and II in Table~\ref{FN}.
\begin{table}[b]
\begin{center}
\begin{tabular}{|c|ccc|ccc|ccc|c|} \hline
 & \bf{10}$_1$ & \bf{10}$_2$ & \bf{10}$_3$ & $\bar{\bf{5}}_1$ & 
$\bar{\bf{5}}_2$ & $\bar{\bf{5}}_3$
& \bf{1}$_1$ & \bf{1}$_2$ & \bf{1}$_3$ & $H$ \\
\hline 
Model I & 2 & 1 & 0 & $\tau+1$ & $\tau$ & $\tau$ & $c$ & $b$ & $a$ & 0
\\
Model II & 3 & 1 & 0 & $\tau$ & $\tau$ & $\tau$ & $c$ & $b$ & $a$ & 0\\
\hline
\end{tabular}
\end{center}
\caption{Froggatt--Nielsen charges for matter and Higgs fields.
In SU(5) language, {\bf{10}}$_i=(Q_i,~\bar{U}_i,~\bar{E}_i)$, 
$\bar{\bf{5}}_i=(\bar{D}_i,~L_i)$, and $H$ is for all Higgses.}
\label{FN}
\end{table}

Mass terms for the lepton sector are given by
\begin{eqnarray}
W &=& \bar{E}_i f_e^{ij} L_j H_d + \bar{N}_i f_{\nu}^{ij} L_j H_u
+\frac{1}{2} \bar{N}_i M_{ij} \bar{N}_j.
\end{eqnarray}
After diagonalizing the charged-lepton Yukawa matrix ($f_e$) and right-handed
neutrino mass matrix ($M$) and taking into account the FN charges, 
we obtain the following mass matrices:
\begin{eqnarray}
m_e &\equiv&\frac{f_e v \cos\beta}{\sqrt{2}}=
 {\rm diag.}(e_1 \epsilon^{3},~e_2 \epsilon,~
e_3 )~ \epsilon^\tau m_{3},
\label{charged_lep_mass}
\\
M &=& {\rm diag.}(n_1 \epsilon^{2c},~n_2 \epsilon^{2b},~
\epsilon^{2a})~ M_R,
\label{right_NR}
\\
m_{\nu D} &\equiv& \frac{f_\nu v \sin\beta}{\sqrt{2}}=
m_3 \epsilon^\tau \left(
\begin{array}{ccc}
\bar{C}_3 \epsilon^{c+\delta} & \bar{B}_3 \epsilon^{c} &
\bar{A}_3 \epsilon^{c}\\
\bar{C}_2 \epsilon^{b+\delta} & \bar{B}_2 \epsilon^{b} &
\bar{A}_2 \epsilon^{b} \\
\bar{C}_1 \epsilon^{a+\delta} & \bar{B}_1 \epsilon^{a} &
\bar{A}_1 \epsilon^{a}
\end{array}
\right),
\label{dirac_neutrino}
\end{eqnarray}
where 
$v\equiv\sqrt{ \langle H_u \rangle^2+ \langle H_d \rangle^2}, 
~\tan\beta\equiv \langle H_u \rangle / \langle H_D \rangle$, and
$m_3$ and $M_R$ represent a weak scale and a right-handed neutrino scale,
respectively.
The coefficients $e_i$, $n_i$, $\bar{A}_i$, $\bar{B}_i$, and 
$\bar{C}_i$ are undetermined but expected to be
of order 1, and $\delta=0$ and 1 for
Models I and II, respectively.
Through the seesaw mechanism, assuming $M_R \gg m_3$, 
tiny neutrino masses can be obtained:
\begin{eqnarray}
m_\nu &=& m_{\nu D}^{\rm T} M^{-1} m_{\nu D},
\nonumber \\
&=& \frac{m_3^2}{M_R}\epsilon^{2\tau}
\left(
\begin{array}{ccc}
C_i^2 \epsilon^{2\delta} & B_i C_i \epsilon^\delta & A_i C_i \epsilon^\delta \\
B_i C_i \epsilon^\delta & B_i^2 & A_i B_i \\
A_i C_i \epsilon^\delta & A_i B_i & A_i^2
\end{array}
\right),
\label{neutrino_mass}
\end{eqnarray}
where coefficients $A_i$, $B_i$, and $C_i$ are given by
\begin{eqnarray}
(A_3,~B_3,~C_3)&=&\frac{1}{\sqrt{n_1}}(\bar{A}_3,~\bar{B}_3,~\bar{C}_3),
\nonumber \\
(A_2,~B_2,~C_2)&=&\frac{1}{\sqrt{n_2}}(\bar{A}_2,~\bar{B}_2,~\bar{C}_2),
\nonumber \\
(A_1,~B_1,~C_1)&=&(\bar{A}_1,~\bar{B}_1,~\bar{C}_1).
\end{eqnarray}
As can be seen in Eq.~(\ref{neutrino_mass}), a large mixing between
$\nu_\mu$ and $\nu_\tau$ can be expected since the matrix elements
$(m_\nu)_{ij}~(i,j=2,3)$ are of the same order.
Note that both Models I and II have the same hierarchical structure
in charged-lepton masses (Eq.~(\ref{charged_lep_mass}));
on the other hand, the neutrino mass matrix Eq.~(\ref{neutrino_mass})
depends on the models ($\delta$). It has been studied in Ref.~\cite{JY} that
in order to obtain the correct masses for charged leptons, the best value
for $\epsilon$ is 0.07. Therefore in our analysis, we will fix $\epsilon$
to 0.07.

The neutrino mass matrix in Eq.~(\ref{neutrino_mass}) is diagonalized
by a mixing matrix $U^{MNS}$:
\begin{eqnarray}
{U^{MNS}}^{\rm T} m_\nu U^{MNS} &=& {\rm diag.}
(m_{\nu 1},~m_{\nu 2},~m_{\nu 3}),\\
\nu_{F \alpha} &=& U^{MNS}_{\alpha i} \nu_{M i},
\end{eqnarray}
where $\nu_{F}$ and $\nu_M$ are the flavor and mass eigenstates of
neutrinos.
The Dirac neutrino mass matrix Eq.~(\ref{dirac_neutrino}) is 
diagonalized by the bi-unitary transformation:
\begin{eqnarray}
{V^{Dirac}}^\dagger m_{\nu D} U^{Dirac} = {\rm diag.}
(m_{\nu D1},~m_{\nu D2},~m_{\nu D3}).
\label{dirac_mixing}
\end{eqnarray}
The mixing matrix $U^{Dirac}$ is relevant to LFV
in the charged lepton sector as we will see later. Although in general 
the matrix $U^{Dirac}$ is different from the neutrino mixing matrix $U^{MNS}$,
an important point is that the large mixing originates from the lopsided
structure in Dirac neutrino masses, and hence both mixing
matrices $U^{MNS}$ and $U^{Dirac}$ possess a large mixing. 
This fact is very important for the LFV phenomena.

We should stress that because of the lopsided FN charges listed in 
Table~\ref{FN}, the Kobayashi--Maskawa matrix in the quark sector
has only small mixings. Therefore, this framework can naturally 
accommodate small mixings in the quark sector as well as large mixing
in the neutrino sector.

\section{Neutrino masses and mixings}
As can be seen in Eq.~(\ref{neutrino_mass}), the mass matrix for neutrinos
does not depend on the FN charges of right-handed neutrinos.
A dependence on the FN charge $\tau$ is found only in the overall factor
of the neutrino mass matrix Eq.~(\ref{neutrino_mass}).
Therefore the predictions for the ratio of neutrino masses and 
the neutrino mixings
are almost the same in the different choices of the charge $\tau$
\footnote{
Running effects in the RG equations depend on
the FN charges of right-handed neutrinos and left-handed leptons. 
Thus the neutrino mass matrix at the low-energy scale can be affected 
by the runnings 
from GUT scale to right-handed neutrino mass scale. However, 
we have checked that these effects are not very significant, so
the results for neutrino masses and mixings are almost
the same in the different cases.}.
However, some of the matrix elements depend on $\delta$. 
In this section, therefore,
we consider Models I and II separately.

\subsection{Model I ($\delta=1$)}
First let us consider Model I.
When we diagonalize the matrix in Eq.~(\ref{neutrino_mass})
with $\delta=1$, we naively obtain
the neutrino mixing matrix $U^{MNS}$ as follows:
\begin{eqnarray}
U^{MNS} &\sim& \left(
\begin{array}{ccc}
1 &  O(\epsilon) & O(\epsilon)
\\
O(\epsilon) & O(1) & O(1)
\\
O(\epsilon) & O(1)  & O(1)
\end{array}
\right),
\label{MNS}
\end{eqnarray}
in the leading order of $\epsilon$.
Since the matrix elements $(m_\nu)_{ij}~(i,j=2,3)$ are of order 1,
we can naturally get a large mixing for atmospheric neutrinos.
For solar neutrinos, it is easy to get a small mixing since $(U^{MNS})_{e2}$
can be naturally small. However, it is not difficult to get
even the large mixing solutions for solar neutrinos 
as we will see in our numerical analysis later.
An interesting point is that a component $U_{e3}^{MNS}$ is expected to be
of order $\epsilon$, so that this is not only consistent with CHOOZ and
atmospheric neutrino experiments, but also can be within the reach of future
neutrino experiments.

In our numerical analysis, 
we randomly generate data sets of coefficients $\bar{A}_i$, 
$\bar{B}_i$, and $\bar{C}_i$,
in which we vary the absolute values of the 
coefficients $\bar{A}_i$, $\bar{B}_i$, and $\bar{C}_i$
from 0.5 to 2 and their phases from 0 to $2\pi$ 
since the coefficients $\bar{A}_i$, $\bar{B}_i$, and $\bar{C}_i$ 
are complex constants of order 1. For simplicity, we fix $n_i$ ($i=1,2$)
to 1 in right-handed neutrino masses in Eq.~(\ref{right_NR}).
Then we calculate neutrino masses and mixings, and look for
neutrino solutions that will solve atmospheric and solar neutrino
problems. In our analysis, we impose the following conditions:
\begin{eqnarray}
\sin^2 2 \theta_{\rm atm} \ge 0.8,
\label{sin_atm}
\end{eqnarray}
for the atmospheric neutrino solution, 
\begin{eqnarray}
&&0.2 \le \tan^2\theta_{\rm sol} \le 1,
\nonumber
\\
&&3.3\times 10^{-3} \le \frac{\delta m^2_{\rm sol}}{\delta m^2_{\rm atm}}
\le 10^{-1},
\label{const_LMA}
\end{eqnarray}
for the large mixing angle (LMA) MSW solution, 
\begin{eqnarray}
&&2 \times 10^{-4} \le \tan^2\theta_{\rm sol} \le 3 \times 10^{-3},
\nonumber
\\
&&1.3 \times 10^{-3} \le \frac{\delta m^2_{\rm sol}}{\delta m^2_{\rm atm}}
\le 3.3 \times 10^{-3},
\label{const_SMA}
\end{eqnarray}
for the small mixing angle (SMA) MSW solution,
\begin{eqnarray}
&&0.3 \le \tan^2\theta_{\rm sol} \le 3,
\nonumber
\\
&&1.3 \times 10^{-7} \le \frac{\delta m^2_{\rm sol}}{\delta m^2_{\rm atm}}
\le 1 \times 10^{-4},
\label{const_LOW}
\end{eqnarray}
for the LOW or vacuum solution.
Here 
$\delta m^2_{\rm atm}=m^2_{\nu 3}-m^2_{\nu 2}$ and
$\delta m^2_{\rm sol}=m^2_{\nu 2}-m^2_{\nu 1}$, and
we define ``effective mixings'' $\sin^2 2\theta_{\rm atm}$
and 
$\tan^2 \theta_{\rm sol}$  as 
\begin{eqnarray}
\sin^2 2\theta_{\rm atm} &\equiv& 4|U_{\mu3}^{MNS}|^2 (1-|U^{MNS}_{\mu3}|^2),
\\
\tan^2 \theta_{\rm sol} &\equiv& |U_{e2}^{MNS}/U_{e1}^{MNS}|^2,
\end{eqnarray}
for three-flavor neutrino oscillation.
We also impose the CHOOZ constraint:
\begin{eqnarray}
|U^{MNS}_{e3}| < 0.15.
\label{const_CHOOZ}
\end{eqnarray}
To fix a scale for neutrino masses, we take the scale for atmospheric
neutrinos as 
\begin{eqnarray}
\delta m^2_{\rm atm} = 3 \times 10^{-3}~~ {\rm eV}^2
\end{eqnarray}
for simplicity.

\begin{figure}
\centerline{
\psfig{figure=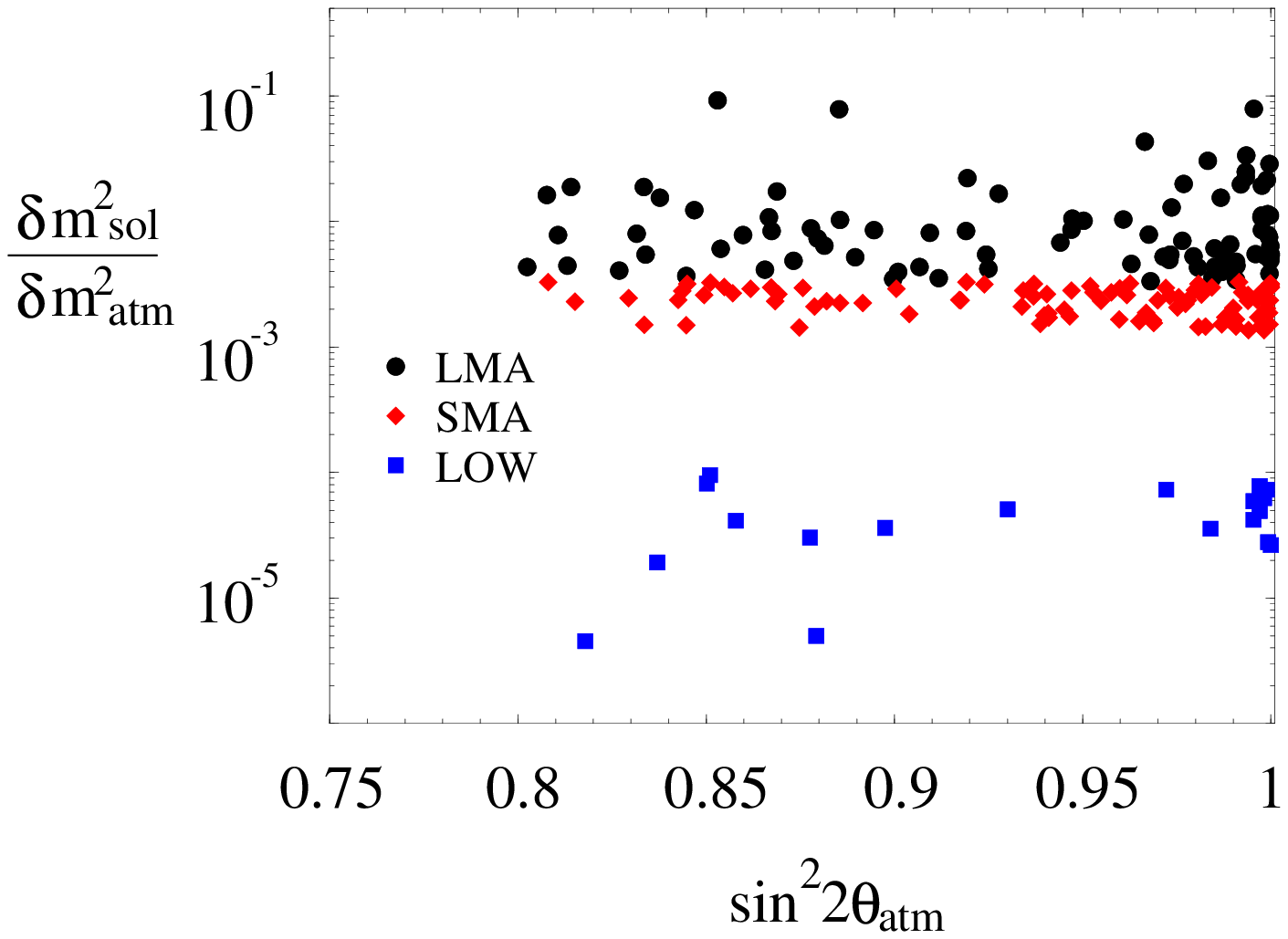,width=0.83\textwidth}
}
\vspace{0.5cm}
\centerline{
\psfig{figure=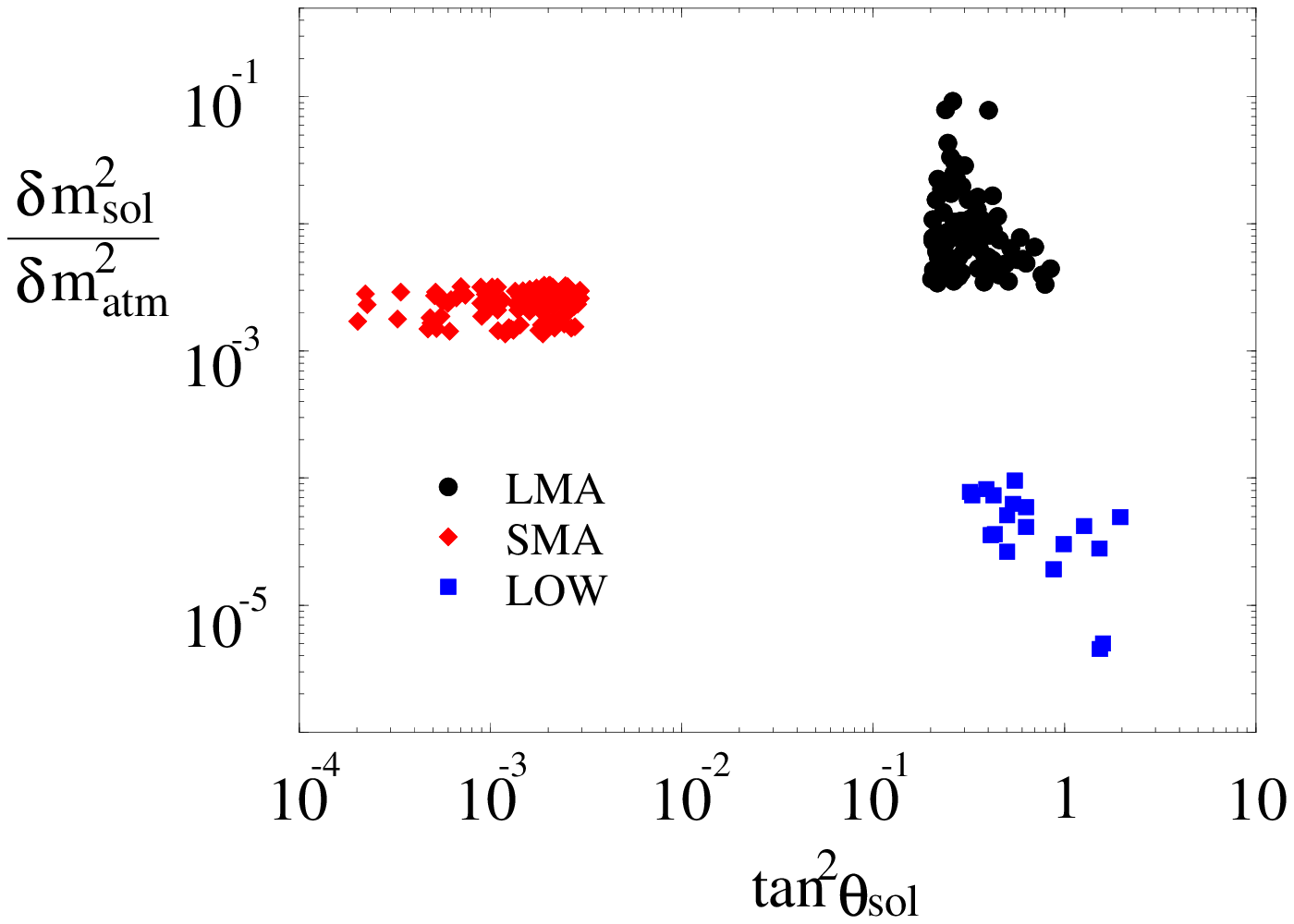,width=0.83\textwidth}
}
\caption{ Distribution of the points of neutrino solutions in Model I.
Black circles satisfy the constraints for the LMA solution,
the diamond-shaped points for the SMA solution, and
the square points for the LOW solution.}
\label{fig_neu}
\end{figure}

With these constraints, $0.1\%$ of data sets out of all data sets we generated
passed the constraints for the CHOOZ, 
atmospheric and SMA solution, $0.3\%$ for the CHOOZ, atmospheric 
and LMA solution, and $0.07\%$ for the CHOOZ, atmospheric and LOW solution.
Therefore this class of models can accommodate various solar neutrino 
solutions (LMA, SMA and LOW), but not the vacuum solution.
In Fig.~{\ref{fig_neu}} we show the distribution of the
neutrino solutions.
All points satisfy the atmospheric neutrino constraint in 
Eq.~(\ref{sin_atm}), as shown in Fig.~\ref{fig_neu}.
The circle points satisfy the condition
for the LMA solution in Eq.~(\ref{const_LMA}),
the diamond-shaped points for the SMA solution in Eq.~(\ref{const_SMA}), 
and the square points for the LOW solution in Eq.~(\ref{const_LOW}).

\begin{figure}
\centerline{
\psfig{figure=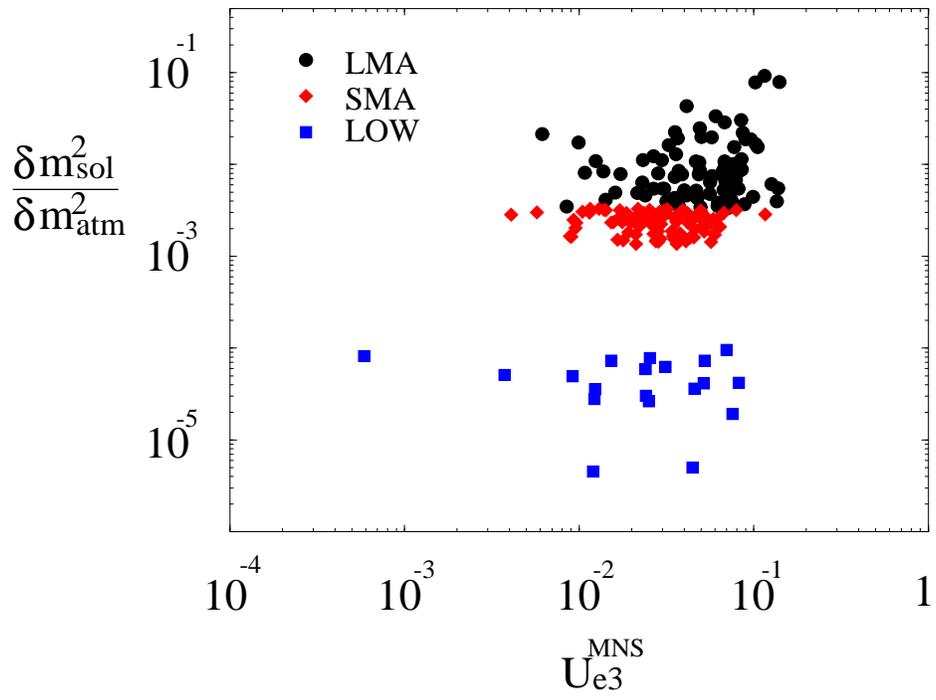,width=0.9\textwidth}
}
\caption{Predicted values of $|U_{e3}^{MNS}|$ in Model I}
\label{ve3}
\end{figure}

One interesting point is that most of the predicted values for 
$|U_{e3}^{MNS}|$ are larger than $10^{-2}$ (Fig.~\ref{ve3}), so that
the values $|U_{e3}^{MNS}|$ can be reached by
future neutrino experiments~\cite{neutrino_factory}. Therefore
the future precise measurement of $|U_{e3}^{MNS}|$ 
will be very important to test this class of models.
\subsection{Model II $(\delta=0)$}
\begin{figure}
\centerline{
\psfig{figure=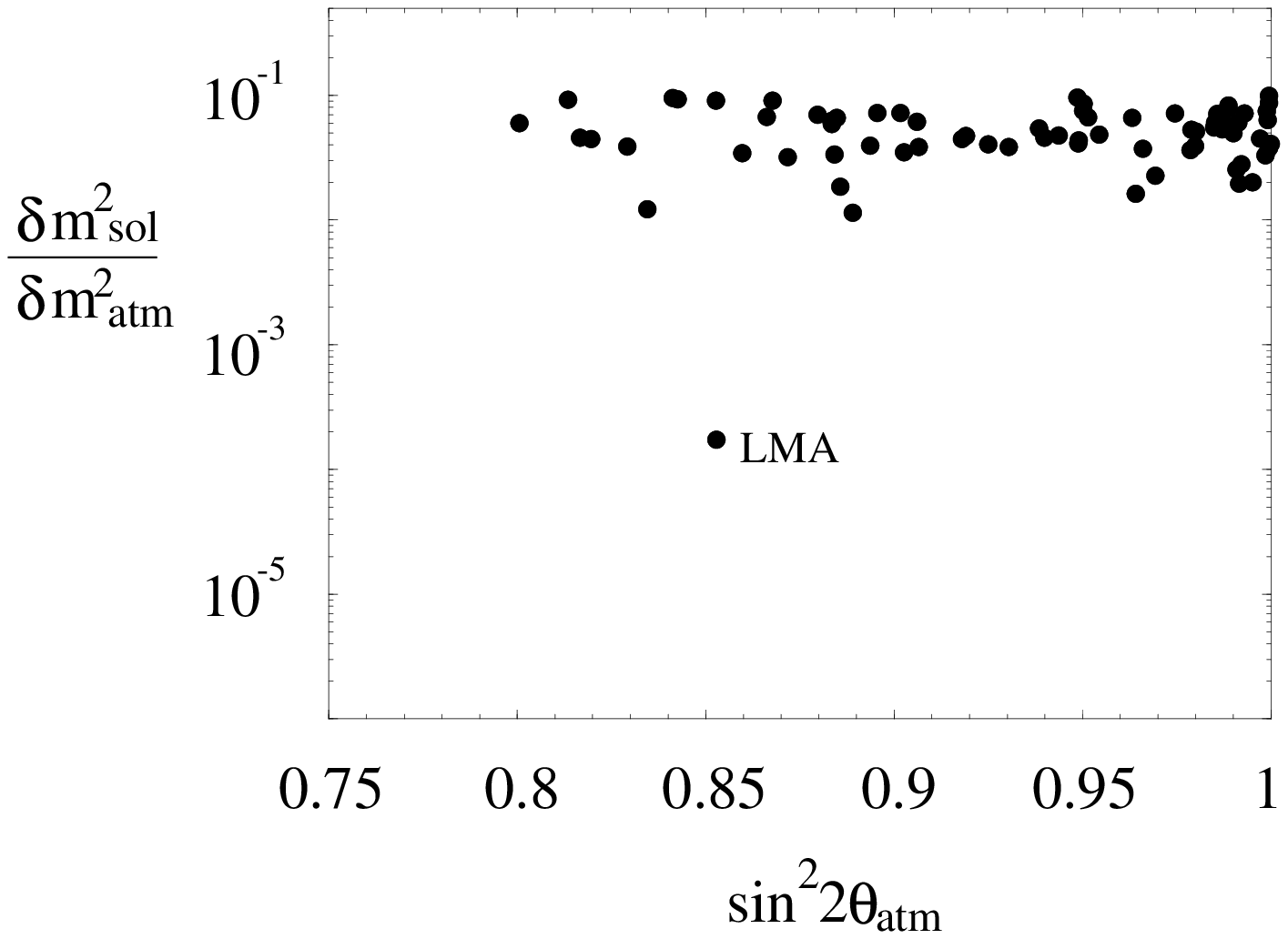,width=0.83\textwidth}
}
\vspace{0.5cm}
\centerline{
\psfig{figure=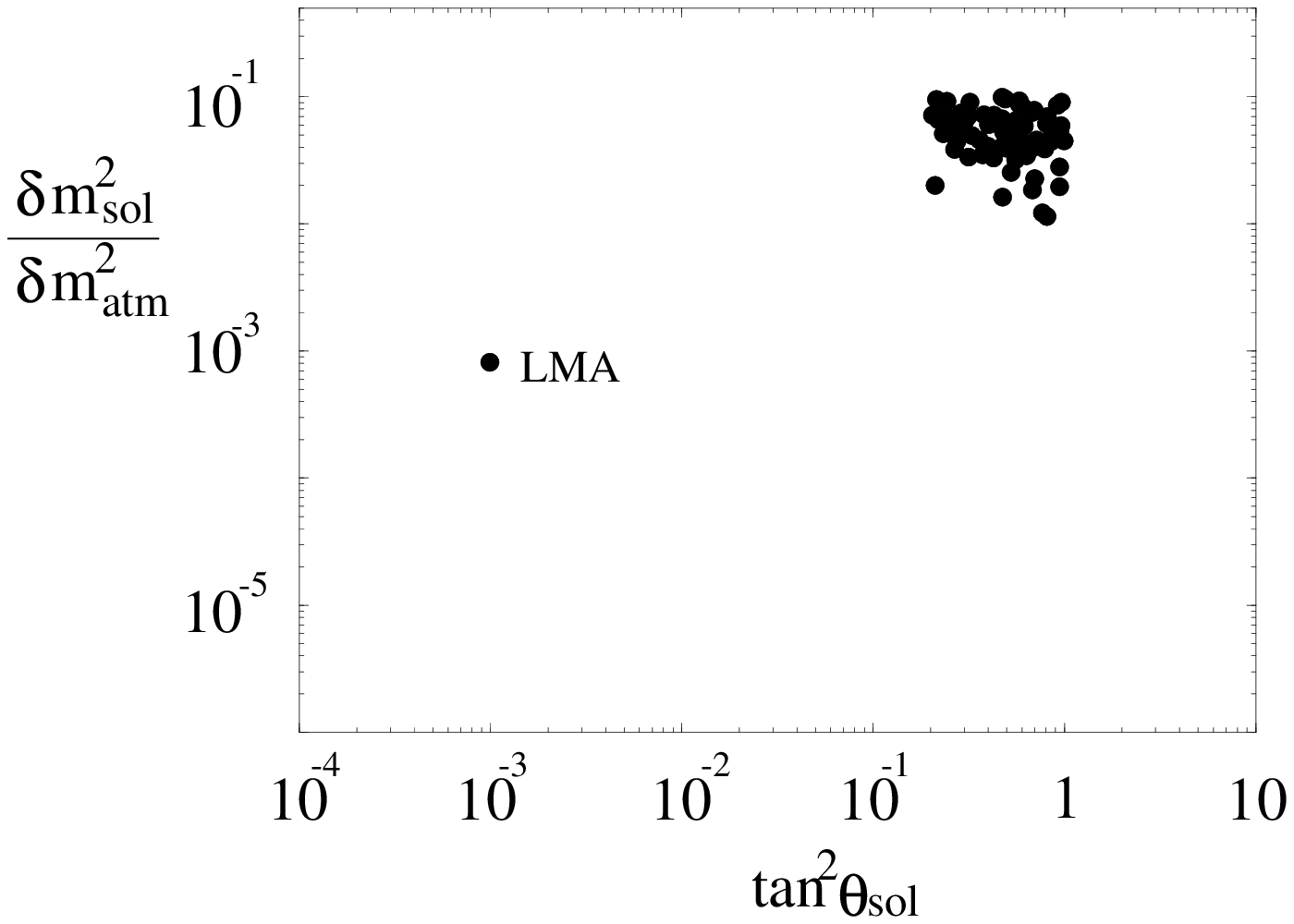,width=0.83\textwidth}
}
\caption{ Distribution of the points of neutrino solutions in Model II.
All points satisfy the constraints for atmospheric and LMA solutions
in Eqs.~(\ref{sin_atm}) and (\ref{const_LMA}).
}
\label{fig_neu_model2}
\end{figure}

\begin{figure}
\centerline{
\psfig{figure=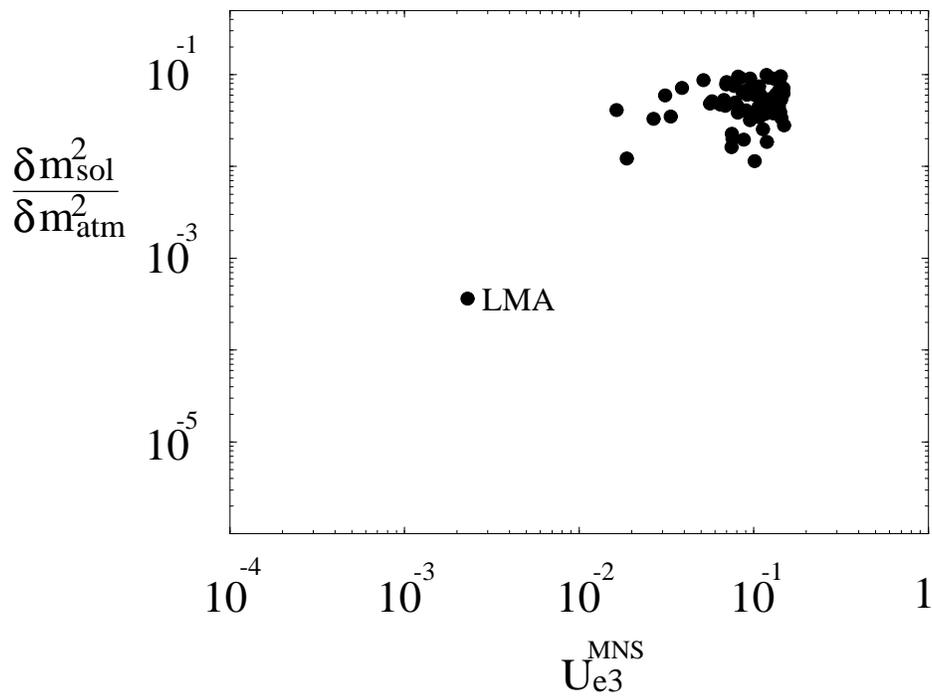,width=0.9\textwidth}
}
\caption{Predicted values of $|U_{e3}^{MNS}|$ in Model II}
\label{ve3_model2}
\end{figure}

In this case, all matrix elements in Eq.~(\ref{neutrino_mass}) with $\delta=0$
are of order 1, and hence we can expect all elements of the neutrino 
mixing matrix to be of order 1: $U^{MNS}_{ij}\sim O(1)~(i,j=1$--3).
Therefore we can naturally get large mixing angles for both atmospheric
and solar neutrinos. 
In our numerical analysis, we impose the same
constraints for atmospheric, solar neutrino solutions and the CHOOZ experiment
as in Model I. Then we search for the solutions. 
In Fig.~\ref{fig_neu_model2},
we show their distribution. As we expected,
the SMA solution is hardly obtained, and it is difficult
to make a large enough mass hierarchy for the LOW solution.
Thus Model II strongly prefers the LMA solution.
$0.01\%$ of the data sets passed the conditions for the LMA solution.
The probability to realize the SMA and LOW solutions is very small.
Therefore a near-future experiment, KamLAND~\cite{KamLAND},
will be able to test this class of models.
In Model II, the rate to realize the LMA solution is much
smaller than in Model I. The main reason is that the limit
on the CHOOZ experiment severely constrains the models, since
the value of $U^{MNS}_{e3}$ tends to become of order 1.
In Fig.~\ref{ve3_model2}, we show the predicted values of 
$U^{MNS}_{e3}$ for Model II.
As can be seen from Fig.~\ref{ve3_model2}, the predicted values
are so large that the future measurement of $U^{MNS}_{e3}$
can observe them~\cite{neutrino_factory}. 
Therefore, again the measurement of $U^{MNS}_{e3}$
is very important.

Another interesting possibility within Model II is that
the inverted hierarchical neutrino masses,
that is $m^2_{\nu 3}\ll m^2_{\nu 1}<m^2_{\nu 2}$
with $\delta m^2_{\rm atm}=m^2_{\nu 3}-m^2_{\nu 2}$ and
$\delta m^2_{\rm sol}=m^2_{\nu 2}-m^2_{\nu 1}$,
could be realized because of
the degeneracy of the neutrino mass matrix elements.
However, we checked that the rate to realize such a possibility
is very small; only $2\times 10^{-3}~\%$ of the data sets passed the criterion
of the inverted hierarchy and the constraints for the LMA solutions.
This possibility can be tested by future neutrino experiments
such as the neutrino factory~\cite{neutrino_factory}, if this could be
realized.

Here we mainly stressed the importance of the measurement of $U^{MNS}_{e3}$.
In addition to this, CP violation and $2\beta 0 \nu$ decay in neutrino
physics are important to see the distinct features for Models I and II,
as pointed out in Ref.~\cite{JY}.

In the next section, using the models that satisfied the neutrino
constraints here, we will discuss in detail the LFV in the 
charged lepton sector.

\section{Lepton-Flavor Violation}
In the presence of non-zero neutrino Yukawa couplings, we can
expect LFV phenomena in the charged-lepton sector.
Within the framework of SUSY models, 
flavor violation in neutrino Yukawa couplings induces 
LFV in slepton masses even if we assume the universal scalar mass for 
all scalars at GUT scale~\cite{BM,HMTYY}.
In the present models, the LFV is generated
in left-handed slepton masses since right-handed neutrinos couple to 
the left-handed lepton multiplets. A RG equation 
for the left-handed doublet slepton masses $(m^2_{\tilde{L}})$ can be 
written as
\begin{eqnarray}
\mu \frac{d(m^2_{\tilde{L}})_{ij}}{d\mu}
&=& \left(\mu \frac{d(m^2_{\tilde{L}})_{ij}}{d\mu} \right)_{\rm MSSM}
\nonumber \\
&&\hspace{-2cm}+\frac{1}{16\pi^2} \left[
m^2_{\tilde{L}} f_\nu^\dagger f_\nu+f_\nu^\dagger f_\nu m^2_{\tilde{L}}
+2(f_\nu^\dagger m^2_{\tilde \nu} f_\nu +\tilde{m}^2_{H_u}
f_\nu^\dagger f_\nu +A_\nu^\dagger A_\nu)\right]_{ij},
\end{eqnarray}
where $m^2_{\tilde{\nu}}$ and $\tilde{m}^2_{H_u}$ are soft SUSY-breaking 
masses for right-handed sneutrinos ($\tilde{\nu}$) and doublet Higgs
($H_u$), respectively.
Here $(\mu d(m^2_{\tilde{L}})_{ij}/{d\mu} )_{\rm MSSM}$ denotes the RG
equation in case of the minimal SUSY SM (MSSM), and the terms explicitly 
written are additional 
contributions in the presence of the neutrino Yukawa couplings.
In a basis where the charged-lepton Yukawa couplings are diagonal, the term
$(\mu d(m^2_{\tilde{L}})_{ij}/{d\mu} )_{\rm MSSM}$ does not provide
any flavor violations. Therefore the only source of LFV comes 
from the additional terms. 
In our analysis, we numerically solve the RG equations, 
and then calculate the event rates for the LFV processes by using the 
complete formula in Ref.~\cite{HMTYY}. 
Here, in order to obtain an approximate estimation for the LFV masses,
let us consider one-iteration approximate solution to the
LFV mass terms ($i\neq j$):
\begin{eqnarray}
(\Delta m^2_{\tilde{L}})_{ij} \simeq -\frac{(6+a_0^2)m_0^2}{16 \pi^2}
(f_\nu^\dagger f_\nu)_{ij} \log\frac{M_G}{M_R}.
\end{eqnarray}
Here we assume a universal scalar mass $(m_0)$ for all scalars and
a universal A-term $(A_f=a_0 m_0 f_f)$ at the GUT scale 
($M_G=2\times 10^{16}$ GeV). 
From Eqs.~(\ref{dirac_neutrino}) and (\ref{dirac_mixing}), 
the solution can be written as
\begin{eqnarray}
(\Delta m^2_{\tilde{L}})_{ij} \simeq -\frac{(6+a_0^2)m_0^2}{16 \pi^2}
U^{Dirac}_{ik} U^{Dirac *}_{jk} |f_{\nu k}|^2 \log\frac{M_G}{M_R}.
\end{eqnarray}
Note that large neutrino Yukawa couplings and large lepton mixings 
$U^{Dirac}_{ij}$ generate the large LFV in the left-handed slepton masses.

The component 
$(\Delta m^2_{\tilde{L}})_{32}$ ($(\Delta m^2_{\tilde{L}})_{21}$)
generates the $\tau \rightarrow \mu \gamma$ 
($\mu \rightarrow e \gamma$) process through diagrams mediated by
sleptons. The decay rates can be 
approximated as follows:
\begin{eqnarray}
{\rm \Gamma} (e_i \rightarrow e_j \gamma) \simeq 
\frac{e^2}{16 \pi} m^5_{e_i} F \left| (\Delta m^2_{\tilde{L}})_{ij} \right|^2.
\label{event_rate}
\end{eqnarray}
Here $F$ is a function of masses and mixings for SUSY particles.
If the non-degeneracy of slepton masses is very small, the function $F$
is approximately process-independent.

Note that the mixing matrix that is relevant to the LFV masses is
$U^{Dirac}$, not the neutrino mixing matrix $U^{MNS}$. The mixing matrix 
$U^{Dirac}$ depends on the FN charges of
right-handed neutrinos $(a,b,c)$ and $\delta$. In the next subsections, 
we discuss the LFV in some interesting cases.

\subsection{Model I ($\delta=1$) with $(a,b,c)=(0,1,2)$}
First we discuss Model I ($\delta=1$) with $(a,b,c)=(0,1,2)$.
In this case, the neutrino Dirac mass matrix in Eq.~(\ref{dirac_neutrino}) is 
given by
\begin{eqnarray}
m_{\nu D} = m_3 \epsilon^\tau \left(
\begin{array}{ccc}
\bar{C}_3 \epsilon^3 & \bar{B}_3 \epsilon^2 & \bar{A}_3 \epsilon^2 \\
\bar{C}_2 \epsilon^2 & \bar{B}_2 \epsilon  & \bar{A}_2 \epsilon \\
\bar{C}_1 \epsilon   & \bar{B}_1  & \bar{A}_1 
\end{array}
\right).
\label{dirac_mass_model1_012}
\end{eqnarray}
When we diagonalize the Dirac mass matrix, we get the following approximate
expression for $U^{Dirac}$:
\begin{eqnarray}
U^{Dirac} &\simeq& \left(
\begin{array}{ccc}
1 & -\frac{\epsilon X}{\sqrt{1+|\bar{B}_1/\bar{A}_1|^2}} & 
\frac{\epsilon \bar{C}_1^*/\bar{A}_1^*}{\sqrt{1+|\bar{B}_1/\bar{A}_1|^2}} \\
\frac{\epsilon \bar{A}_2 \bar{C}_1}{\bar{A}_1 \bar{B}_2 - \bar{B}_1\bar{A}_2} &
\frac{1}{\sqrt{1+|\bar{B}_1/\bar{A}_1|^2}} & 
\frac{\bar{B}_1^*/\bar{A}_1^*}{\sqrt{1+|\bar{B}_1/\bar{A}_1|^2} }
\\
-\frac{\epsilon \bar{B}_2 \bar{C}_1 }{\bar{A}_1\bar{B}_2 - 
\bar{B}_1\bar{A}_2} &
-\frac{\bar{B}_1/\bar{A}_1}
{\sqrt{1+|\bar{B}_1/\bar{A}_1|^2}} & \frac{1}{\sqrt{1+|\bar{B}_1/\bar{A}_1|^2}}
\end{array}
\right),
\label{dirac_mix_matrix}
\end{eqnarray}
where $X=\bar{C}_1^*(\bar{A}_1 \bar{A}_2^*+\bar{B}_1 \bar{B}_2^*)/
\bar{A_1}(\bar{A}_1^* \bar{B}_2^*-\bar{A}_2^* \bar{B}_1^*)$. 
This expression is a leading order
in terms of $\epsilon$. 
An important point is that the lepton mixing matrix $U^{Dirac}$ 
in Eq.~(\ref{dirac_mix_matrix}) also 
contains a large mixing because of the lopsided
structure of the Dirac neutrino masses. 

\begin{figure}
\vspace*{-0.5cm}
\centerline{
\psfig{figure=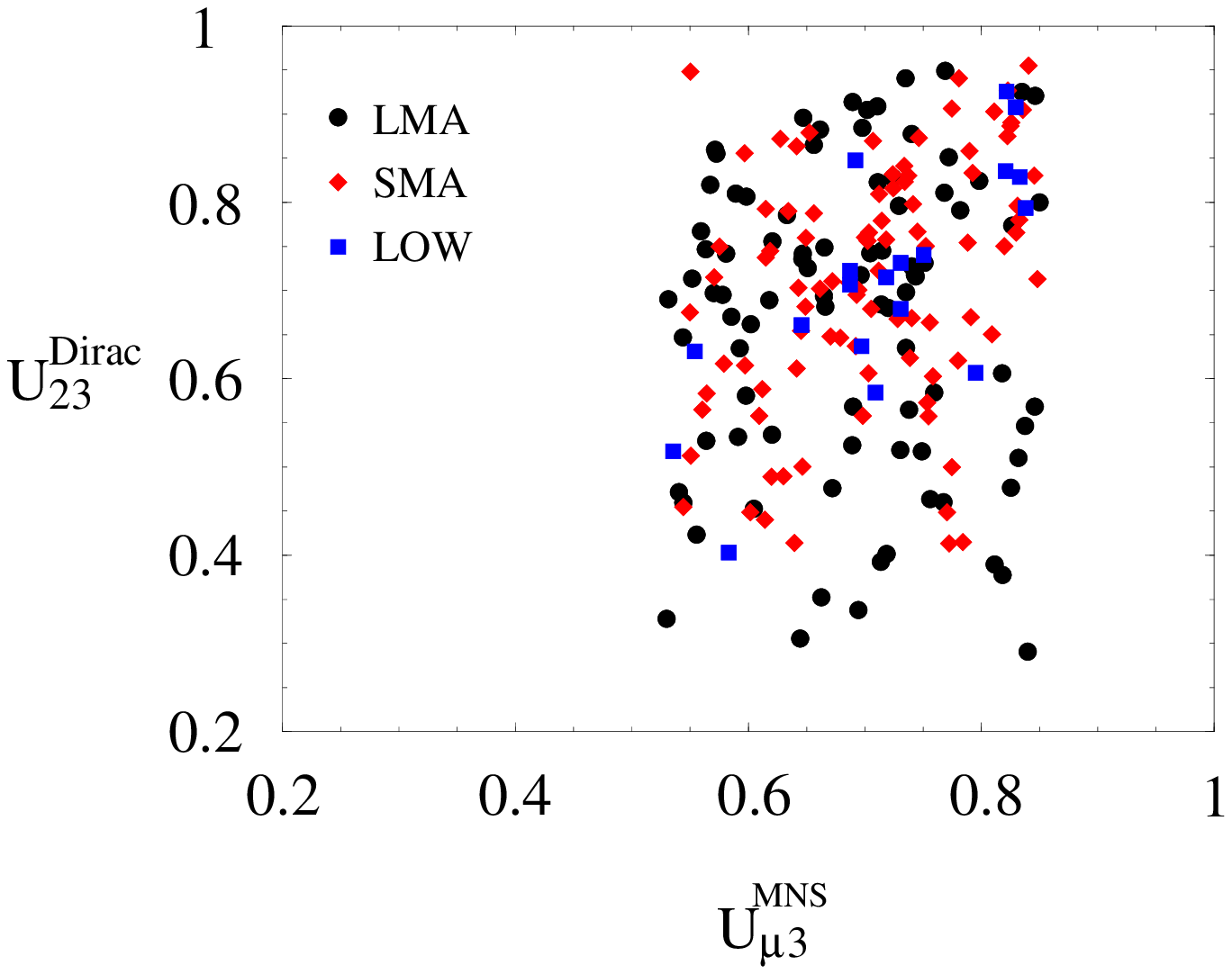,width=0.8\textwidth}
}
\centerline{
\psfig{figure=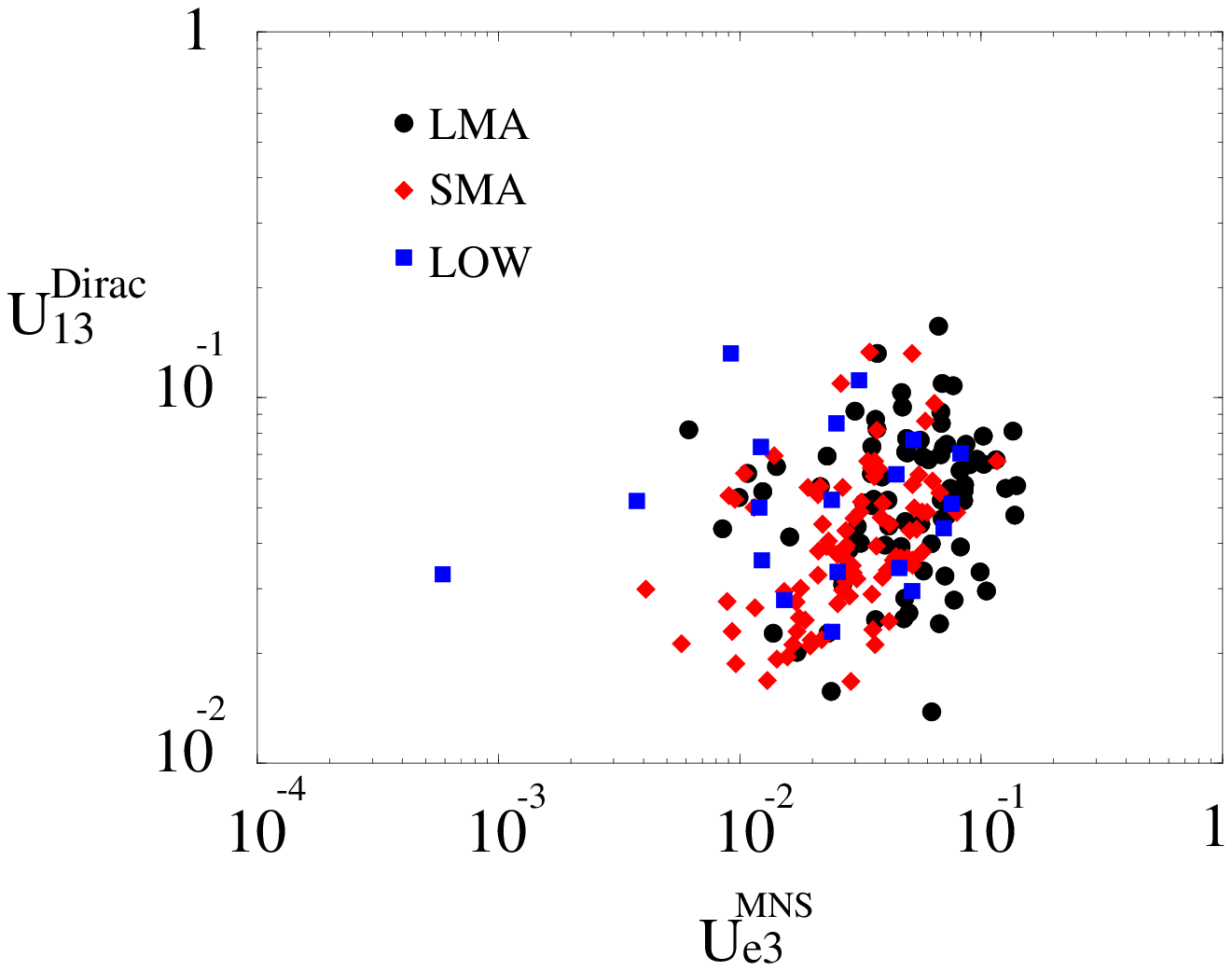,width=0.8\textwidth}
}
\caption{$U^{MNS}$ versus $U^{Dirac}$ in Model I
with $(a,b,c)=(0,1,2)$.}
\label{Vd_Vn_13}
\end{figure}

The component $(\Delta m^2_{\tilde{L}})_{32}$ induces the
$\tau \rightarrow \mu \gamma$ process.
Taking into account the hierarchical structure of neutrino Yukawa couplings
($f_{\nu 1}:f_{\nu 2}:f_{\nu 3}\sim \epsilon^2:\epsilon:1$), 
a dominant term of the matrix element $(\Delta m^2_{\tilde{L}})_{32}$ 
can be written as
\begin{eqnarray}
(\Delta m^2_{\tilde{L}})_{32} &\simeq& -\frac{(6+a_0^2)m^2_0}
{16 \pi^2} U_{33}^{Dirac} U_{23}^{Dirac *} |f_{\nu 3}|^2 
\log \frac{M_G}{M_R}.
\label{32_comp}
\end{eqnarray}
The component $(\Delta m^2_{\tilde{L}})_{21}$, on the other hand, 
generates the $\mu \rightarrow e \gamma$ process. Since $U_{13}^{Dirac}$
is of order $\epsilon$, a leading term of $(\Delta m^2_{\tilde{L}})_{21}$
can be written as
\begin{eqnarray}
(\Delta m^2_{\tilde{L}})_{21} &\simeq& -\frac{(6+a_0^2)m^2_0}
{16 \pi^2} U_{23}^{Dirac} U_{13}^{Dirac *} |f_{\nu 3}|^2 
\log \frac{M_G}{M_R}.
\label{21_comp}
\end{eqnarray}
The branching ratio for the $\tau \rightarrow \mu \gamma$ 
($\mu \rightarrow e\gamma$)
process is proportional to the square of $(\Delta m^2_{\tilde{L}})_{32}$
($(\Delta m^2_{\tilde{L}})_{21}$), as given in Eq.~(\ref{event_rate}). 
Thus the important parameters for the branching ratios are
the matrix elements $U^{Dirac}_{23}$, $U^{Dirac}_{13}$ and $U^{Dirac}_{33}$,
and the third-generation Yukawa coupling $f_{\nu 3}$.
In Model I, the element $U^{Dirac}_{23}$ is expected to be of
order 1 because of the lopsided structure in the Dirac mass terms as 
expressed in Eq.~(\ref{dirac_mix_matrix}).
In Fig.~\ref{Vd_Vn_13}, we show the numerical result for $U^{Dirac}_{23}$
compared to $U^{MNS}_{\mu 3}$. As expected from 
Eq.~(\ref{dirac_mix_matrix}), $U^{Dirac}_{13}$ is of the order of $\epsilon$,
as shown in Fig.~\ref{Vd_Vn_13}.
The order-1 $U^{Dirac}_{23}$ and the order-$\epsilon$ (non-zero)
$U^{Dirac}_{13}$ can induce a significantly large event rate for
$\mu \rightarrow e \gamma$ process if the neutrino Yukawa coupling
$f_{\nu 3}$ is large, as pointed out in Ref.~\cite{JTY}.

A magnitude of $f_{\nu 3}$ depends on the FN charge $\tau$. Even at this
stage, however, we can estimate the ratio of
the branching ratios of the $\mu \rightarrow e \gamma$ and $\tau \rightarrow 
\mu \gamma$ processes as ${\rm Br}(\mu \rightarrow e \gamma)/
{\rm Br}(\tau \rightarrow \mu \gamma)$, which is $\tau$-independent. 
If we consider a value of
$\Gamma(e_i \rightarrow e_j \gamma)/\Gamma(e_i \rightarrow e_j \nu_i
\bar{\nu}_j)$, the dependence of the initial lepton mass in the process 
is cancelled. Thus
we can expect the following relation from Eqs.~({\ref{event_rate}}),
(\ref{32_comp}), and (\ref{21_comp}):
\begin{eqnarray}
\frac
{\Gamma(\mu \rightarrow e \gamma)/
\Gamma(\mu \rightarrow e \nu_\mu \bar{\nu_e})}
{\Gamma(\tau \rightarrow \mu \gamma)/
\Gamma(\tau \rightarrow \mu \nu_\tau \bar{\nu_\mu})}
&\sim& \left| \frac{(\Delta m^2_{\tilde{L}})_{21}}
{(\Delta m^2_{\tilde{L}})_{32}} \right|^2 
\simeq \left|\frac{U^{Dirac}_{13}}{U^{Dirac}_{33}} \right|^2
\sim \epsilon^2\left|\frac{\bar{C}_1}{\bar{A}_1}\right|^2.
\end{eqnarray}
Taking into account ${\rm Br}(\mu \rightarrow e \nu_\mu\bar{\nu_e}) 
\simeq 100\%$
and ${\rm Br}(\tau \rightarrow \mu \nu_\tau \bar{\nu_\mu} )\simeq 17\%$,
we obtain
\begin{eqnarray}
\frac{{\rm Br}(\mu \rightarrow e \gamma)}
{{\rm Br}(\tau \rightarrow \mu \gamma)} &\sim& 
\frac{\epsilon^2}{0.17}\left|\frac{\bar{C}_1}{\bar{A}_1}\right|^2
\sim 0.03 \left(\frac{\epsilon}{0.07}\right)^2 
\left(\frac{|\bar{C}_1/\bar{A}_1|}{1.0}\right)^2.
\label{predic}
\end{eqnarray}
This relation is a $\tau$-independent prediction of Model I
with $(a,b,c)=(0,1,2)$.
Since the current experimental limits on these processes
are ${\rm Br}(\mu \rightarrow e \gamma)<1.2 \times 10^{-11}$
and ${\rm Br}(\tau \rightarrow \mu \gamma)<1.1 \times 10^{-6}$~\cite{PDG},
the process $\mu \rightarrow e \gamma$ is much more sensitive to this 
class of models.

In order to discuss the absolute values of the branching ratios
for LFV processes, we need to fix the FN charge parameter $\tau$.
In SUSY models, there are two interesting cases.
In the models with $\tau=0$, it is suggested that all the third-generation 
Yukawa couplings are of order 1. In SUSY models, this is realized in
the case with the large $\tan \beta$ ($\tan \beta\simeq 50$).
In the models with $\tau=1$, all the third-generation Yukawa couplings
are of order $\epsilon$,
except the top Yukawa coupling, which is expected to be of order 1.
This is likely to be the case with small $\tan \beta$
($\tan \beta\simeq 5$) in SUSY models.
Therefore, in the next subsections, we will consider the two cases,
Case 1, $\tau=0$ and Case 2, $\tau=1$, to see how large the predicted
branching ratios for LFV processes can be.

\subsubsection{Case 1, $\tau=0$}
In Case 1, the models suggest that all third-generation
Yukawa couplings at the GUT scale are of order 1. Therefore,
a large $\tan\beta$ is preferred. Here we take $\tan \beta=50$.
In our numerical analysis,
we impose $m_3 =m_{\rm top}$ at GUT scale in the Dirac neutrino 
mass Eq.~(\ref{dirac_mass_model1_012}) with $\tau=0$,
and then we require the neutrino mass squared difference
to be $\delta m^2_{32} =3\times 10^{-3}$ eV$^2$ 
in order to fix the right-handed neutrino mass scale $M_R$.
We have checked that all third-generation Yukawa couplings
are of the same order, which is about 0.5--0.8.
Since the third-generation neutrino Yukawa coupling is
as large as the top Yukawa coupling, so the large LFV masses are generated 
through the RGE running. Furthermore,
since $\tan \beta$ is very large, the low-energy amplitudes of LFV
processes are enhanced. Therefore, we can expect very large branching
ratios for LFV processes~\cite{JTY}.
\begin{figure}
\vspace*{-0.5cm}
\centerline{
\psfig{figure=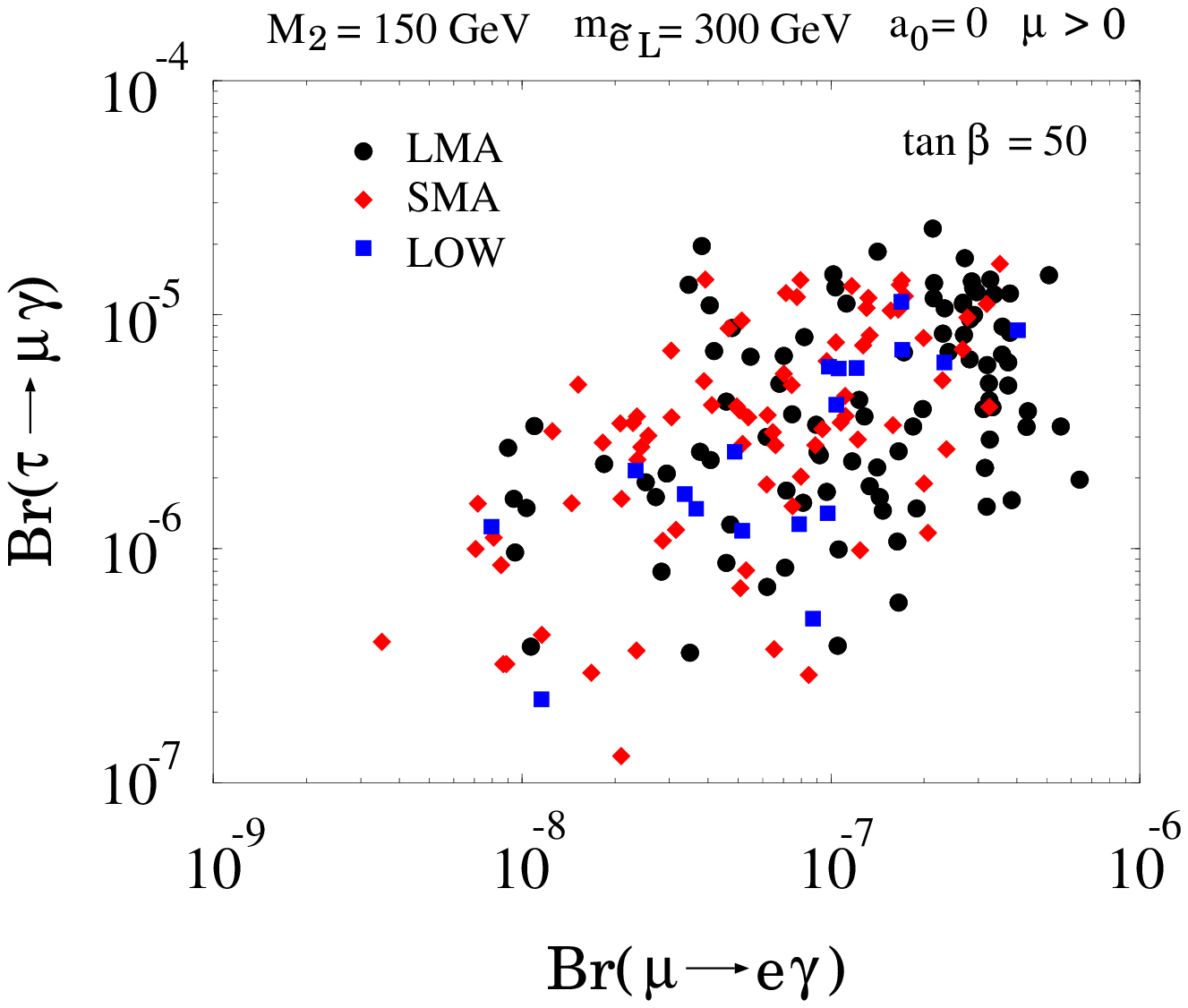,width=0.75\textwidth}
}
\caption{ Br($\mu \rightarrow e \gamma$) versus 
Br($\tau \rightarrow \mu \gamma$) in Model I with $(a,b,c,\tau)=(0,1,2,0)$.
Here we take the left-handed slepton mass to be 300 GeV, the Wino mass
to be 150 GeV, and $\epsilon=0.07$.}
\label{lfv_tan50}
\centerline{
\psfig{figure=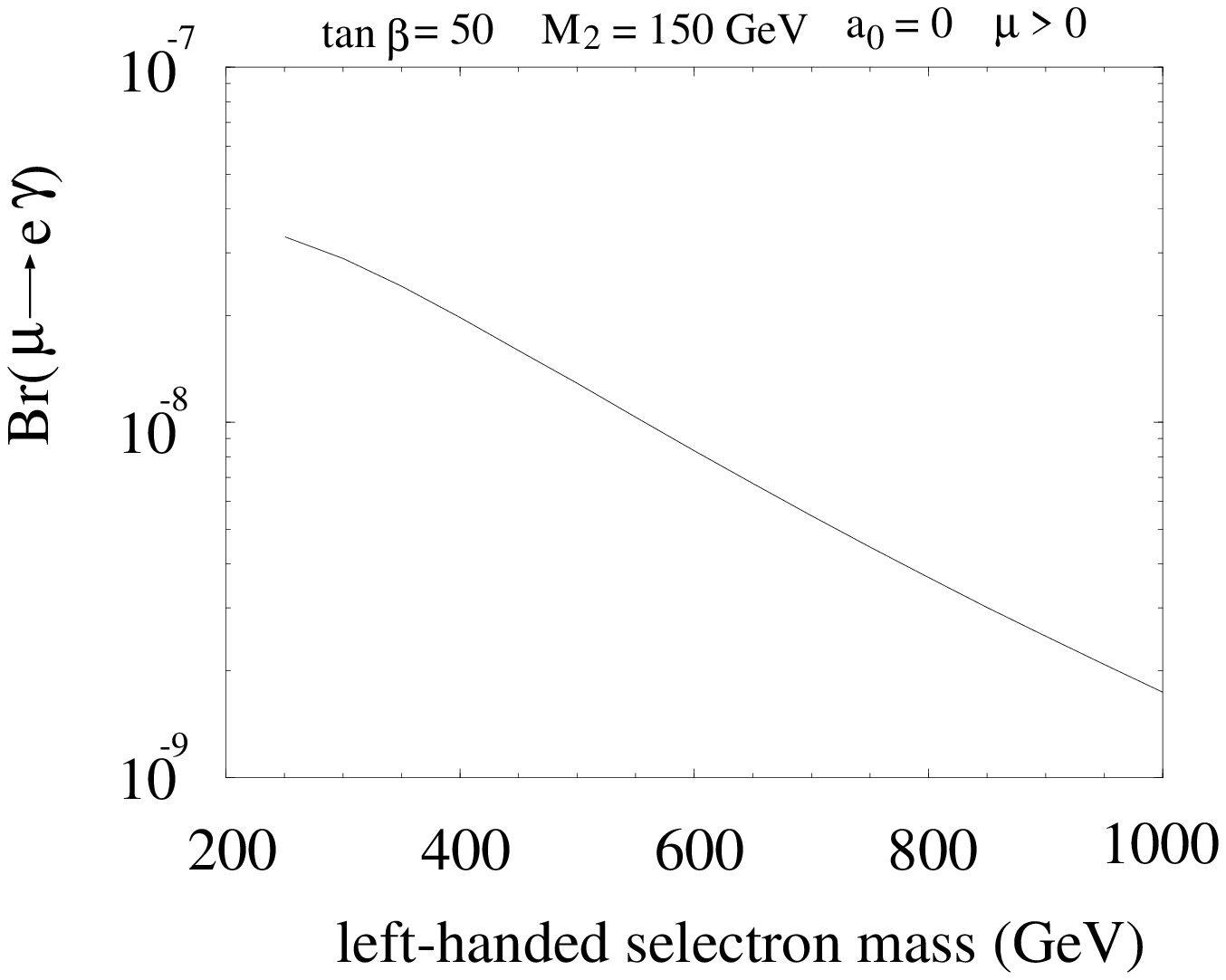,width=0.75\textwidth}
}
\caption{ Br($\mu \rightarrow e \gamma$) as a function of
the left-handed selectron mass in Model I with $(a,b,c)=(0,1,2,0)$.}
\label{meg_SUSY}
\end{figure}
In Fig.~\ref{lfv_tan50}, we show the numerical result for
Br($\tau \rightarrow \mu \gamma$)
and Br($\mu \rightarrow e \gamma$).
Here the circle points satisfy the LMA solution for solar neutrinos,
the diamond-shaped points the SMA solution, and the square points the LOW
solution.
As can be seen from the figure, the branching ratios for the LFV processes
do not depend on the solar neutrino solutions since the important parameters
for $\mu \rightarrow e \gamma$ and $\tau \rightarrow \mu \gamma$ branching 
ratios are only $U^{Dirac}_{13}$, $U^{Dirac}_{23}$ and $U^{Dirac}_{33}$,
which are not affected very much by the constraints on solar neutrino 
solutions, as shown in Fig.~\ref{Vd_Vn_13}.
We also see that the estimation Eq.~(\ref{predic})
is approximately correct, so that the limit from the $\mu \rightarrow e \gamma$
process can provide much stronger constraints on the models.
In Fig.~\ref{meg_SUSY}, we select one point from Fig.~\ref{lfv_tan50}
and show the slepton mass dependence of the branching ratio 
for the $\mu \rightarrow e \gamma$ process.
In this case, these branching ratios are too large to be consistent with 
the present experimental bounds. Therefore the large region of parameter 
space is already excluded.

These results illustrate the significant 
potential of the LFV searches to probe the realistic neutrino models.

\subsubsection{Case 2, $\tau=1$}
\begin{figure}
\vspace*{-0.5cm}
\centerline{
\psfig{figure=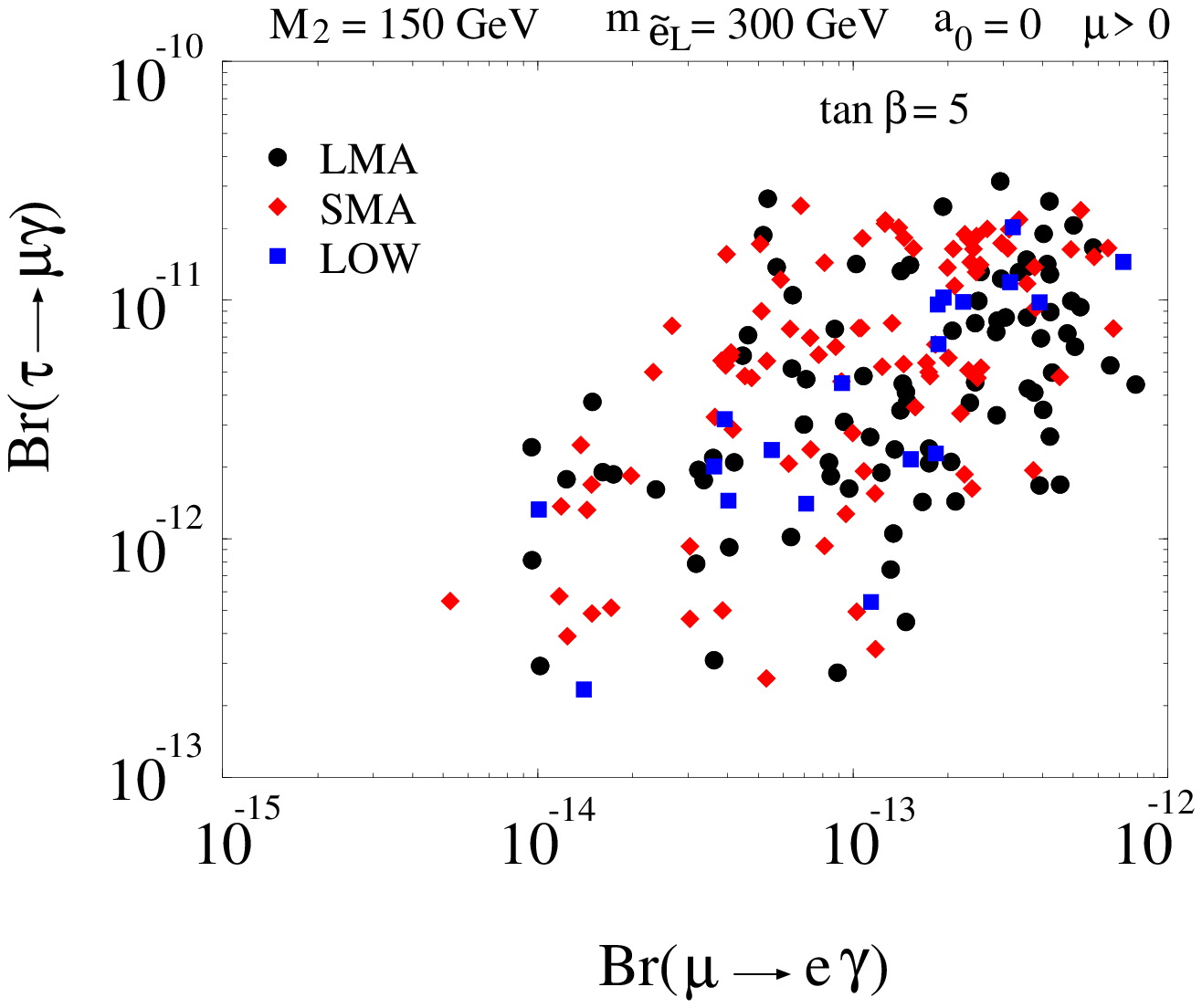,width=0.7\textwidth}
}
\caption{Br($\mu \rightarrow e \gamma$) versus
Br($\tau \rightarrow \mu \gamma$) in Model I with $(a,b,c,\tau)=(0,1,2,1)$.
Here we take the left-handed slepton mass to be 300 GeV, the Wino mass
to be 150 GeV, and $\epsilon=0.07$.}
\label{lfv_tan5}
\centerline{
\psfig{figure=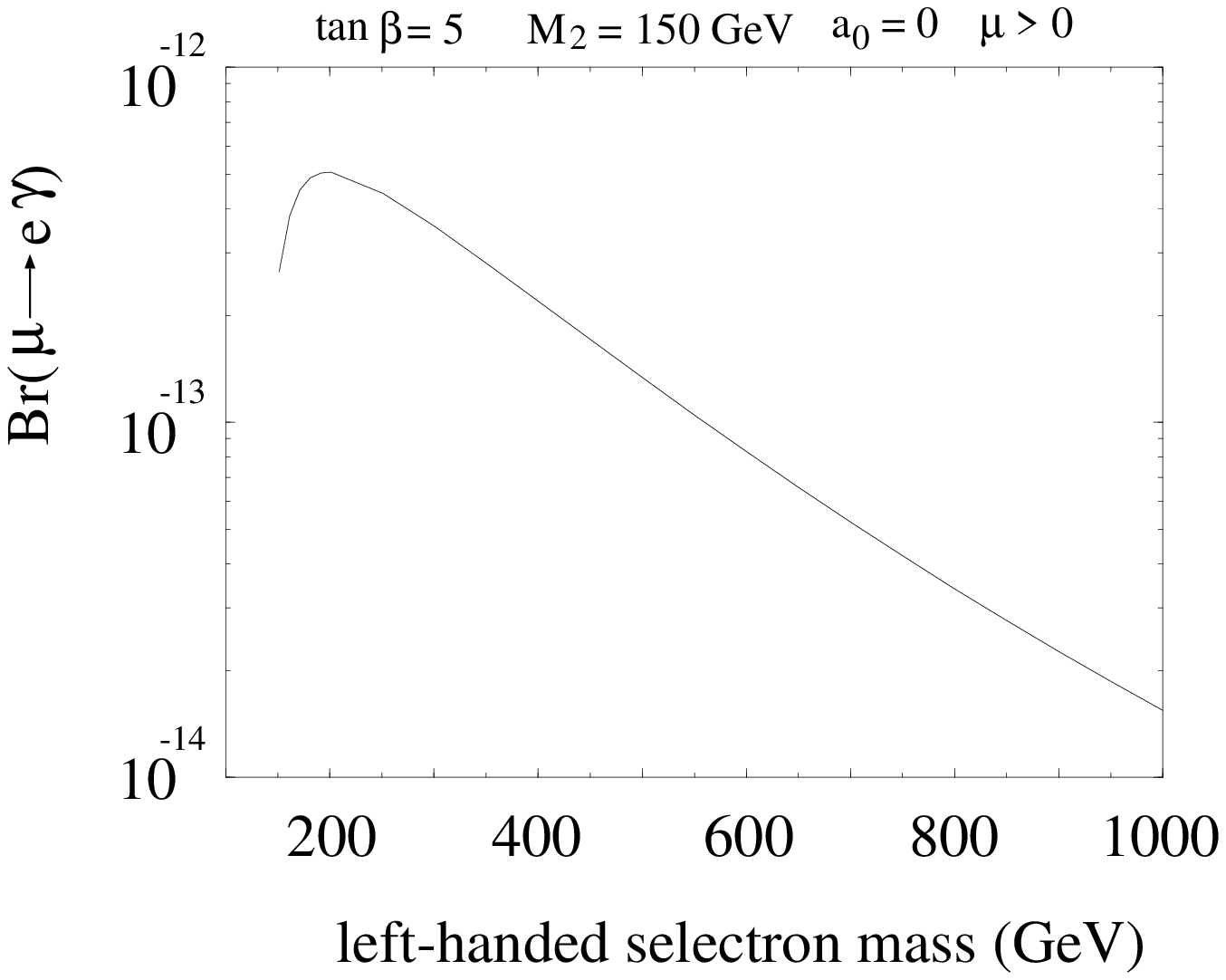,width=0.75\textwidth}
}
\caption{Br($\mu \rightarrow e \gamma$) as a function of
the left-handed selectron mass in Model I with $(a,b,c,\tau)=(0,1,2,1)$.}
\label{meg_SUSY_tan5}
\end{figure}

In models with $\tau = 1$, the top Yukawa coupling is expected to be
of order 1, while the other third-generation Yukawa 
couplings (bottom, tau and tau-neutrino Yukawa couplings)
are of order $\epsilon$. It is thus
likely to be the case with relatively small $\tan \beta$.
Here we take $\tan \beta=5$, and we impose $m_3= m_{\rm top}$ at GUT scale
in the Dirac neutrino mass matrix in Eq.~(\ref{dirac_mass_model1_012}) 
with $\tau=1$, in other words, 
$(f_{\nu})_{33}=\epsilon f_{top} \bar{A}_1$.
We checked that in the case with $\tan \beta=5$,
the bottom and tau Yukawa couplings approximately satisfy the condition
$f_b \simeq f_\tau \simeq \epsilon f_{top}$.
\footnote{The $\tan \beta$-dependence
of the branching ratios is approximately $\tan^2\beta$.}
Therefore, as compared with the case with $\tau=0$, 
the LFV masses in Eqs.~(\ref{32_comp}) and (\ref{21_comp}) 
are suppressed by $\epsilon^2$, and then the branching
ratios for the $\tau \rightarrow \mu \gamma$ and $\mu \rightarrow e \gamma$
processes are reduced by $\epsilon^4$.
The numerical results for the branching ratios 
are shown in Fig.~\ref{lfv_tan5}.
Note that the predicted branching ratios in all solar neutrino solutions
are almost the same, and that the relation in Eq.~(\ref{predic}) is 
approximately
satisfied. Thus the $\mu \rightarrow e \gamma$ search is much more sensitive
to this class of models, unless the future sensitivity of 
Br($\tau\rightarrow \mu \gamma$) can reach values much below $10^{-10}$.
In Fig.~\ref{meg_SUSY_tan5} we also show the branching ratio for 
$\mu \rightarrow e \gamma$ as a function of the left-handed selectron mass.
As can be seen from Figs.~\ref{lfv_tan5} and \ref{meg_SUSY_tan5},
it is very interesting that the predicted branching ratios for 
$\mu \rightarrow e \gamma$ can be just below the present experimental bound
(Br$(\mu \rightarrow e \gamma)<1.2 \times 10^{-11}$~\cite{PDG})
in a wide range of parameter space.
We also get the ratios between branching ratios 
Br($\mu \rightarrow eee$)/Br($\mu \rightarrow e\gamma$)
and R($\mu \rightarrow e$ in Ti (Al))/Br($\mu \rightarrow e\gamma$) as follows:
\begin{eqnarray}
\frac{{\rm Br}(\mu \rightarrow eee)}{{\rm Br}(\mu \rightarrow e\gamma)}
&\simeq& 6\times 10^{-3},
\label{mu_3e}
\\
\frac{{\rm R}(\mu \rightarrow e ~{\rm in~ Ti~(Al)})}
{{\rm Br}(\mu \rightarrow e\gamma)}
&\simeq& 5~(3)\times 10^{-3}.
\label{mu_e_conv}
\end{eqnarray}
In the present models, these ratios are quite predictive 
since the on-shell photon 
penguin diagram, which induces the $\mu \rightarrow e \gamma$ process, 
dominates over the other contributions to the 
$\mu \rightarrow eee$ and $\mu \rightarrow e$ conversion processes.
Therefore the future improvements of the branching ratios,
Br($\mu \rightarrow e \gamma )\sim 10^{-14}$~\cite{PSI},
and R($\mu \rightarrow e$ in Al$)\sim 10^{-16}$~\cite{MECO}
(and possibly R($\mu \rightarrow e$ in Ti$)\sim 10^{-18}$~\cite{PRISM})
will provide a significant impact to this class of models.

\subsection{Model I with $(a,b,c)=(0,0,0)$}
In this case, the Dirac neutrino mass matrix in Eq.~(\ref{dirac_neutrino})
is given by 
\begin{eqnarray}
m_{\nu D} = m_3 \epsilon^\tau \left(
\begin{array}{ccc}
\bar{C}_3 \epsilon & \bar{B}_3  & \bar{A}_3 \\
\bar{C}_2 \epsilon & \bar{B}_2  & \bar{A}_2 \\
\bar{C}_1 \epsilon & \bar{B}_1  & \bar{A}_1 
\end{array}
\right).
\label{dirac_mass_model1_000}
\end{eqnarray}
Then we obtain the mixing matrix $U^{Dirac}$ as follows:
\begin{eqnarray}
U^{Dirac} &\simeq& \left(
\begin{array}{ccc}
1 & \epsilon \frac{X_-}{\sqrt{N_-}} & \epsilon \frac{X_+}{\sqrt{N_+}}\\
\epsilon Y_0 & \frac{Y_-}{\sqrt{N_-}} & \frac{Y_+}{\sqrt{N_+}} \\
\epsilon Z_0 & \frac{Z}{\sqrt{N_-}} & \frac{Z}{\sqrt{N_+}}
\end{array}
\right),
\label{Vdirac_000}
\end{eqnarray}
in the leading order of $\epsilon$. Here the coefficients $X_{\pm}$,
$Y_{\pm (0)}$, $Z_{(0)}$, and $N_{\pm}$ are expressed by 
\begin{eqnarray}
Y_0 &=& \frac{(\bar{A}_i \bar{B}_i^*)(\bar{A}_j^* \bar{C}_j)
-(\bar{A}_i \bar{A}_i^*)(\bar{B}_j^* \bar{C}_j)}
{(\bar{A}_i^* \bar{A}_i)(\bar{B}_j^* \bar{B}_j)-(\bar{A}_i \bar{B}^*_i)
(\bar{A}_j^* \bar{B}_j)},~~
Z_0 = \frac{(\bar{A}_i^* \bar{B}_i)(\bar{B}_j^* \bar{C}_j)-
(\bar{A}_i^* \bar{C}_i)(\bar{B}_j^* \bar{B}_j)}
{(\bar{A}_i^* \bar{A}_i)(\bar{B}_j^* \bar{B}_j)-(\bar{A}_i \bar{B}^*_i)
(\bar{A}_j^* \bar{B}_j)},
\nonumber 
\\
Y_\pm &=& \frac{-(\bar{A}_i^* \bar{A}_i)+(\bar{B}_i^* \bar{B}_i)
\pm \sqrt{\left\{ (\bar{A}_i^* \bar{A}_i)-(\bar{B}_i^* \bar{B}_i) \right\}^2 
+4(\bar{A}_i^* \bar{B}_i)(\bar{A}_j \bar{B}_j^*)}}{2},
\nonumber
\\
Z &=& (\bar{A}_i^* \bar{B}_i),~~
X_\pm = -Y_\pm Y_0^*-Z_\pm Z_0^*,~~
N_\pm = |Y_\pm|^2+|Z_\pm|^2.
\label{Vdirac_000_no2}
\end{eqnarray}
Here summation over $i$ and $j$ $(i,j=1,2)$ is assumed: for example, 
$(\bar{A}_i^* \bar{B}_i) \equiv \sum_{i=1-2} \bar{A}_i^* \bar{B}_i$.
The main difference from that in the previous case (see 
Eq.~(\ref{dirac_mix_matrix})) is that the matrix elements $U^{Dirac}_{ij}$
depend on more parameters, since the matrix elements $(m_{\nu D})_{ij}~
(i=1,2,~j=1-3)$ are larger than those in Eq.~(\ref{dirac_mass_model1_012}).
However, the important point is the same, that is
the matrix elements $U^{Dirac}_{ij}$ $(i,j=2,3)$ in 
Eq.~(\ref{Vdirac_000}) are of order 1 because of the lopsided 
structure of the Dirac neutrino mass matrix, and the element 
$U^{Dirac}_{13}$ is of order $\epsilon$.

\begin{figure}
\vspace*{-0.5cm}
\centerline{
\psfig{figure=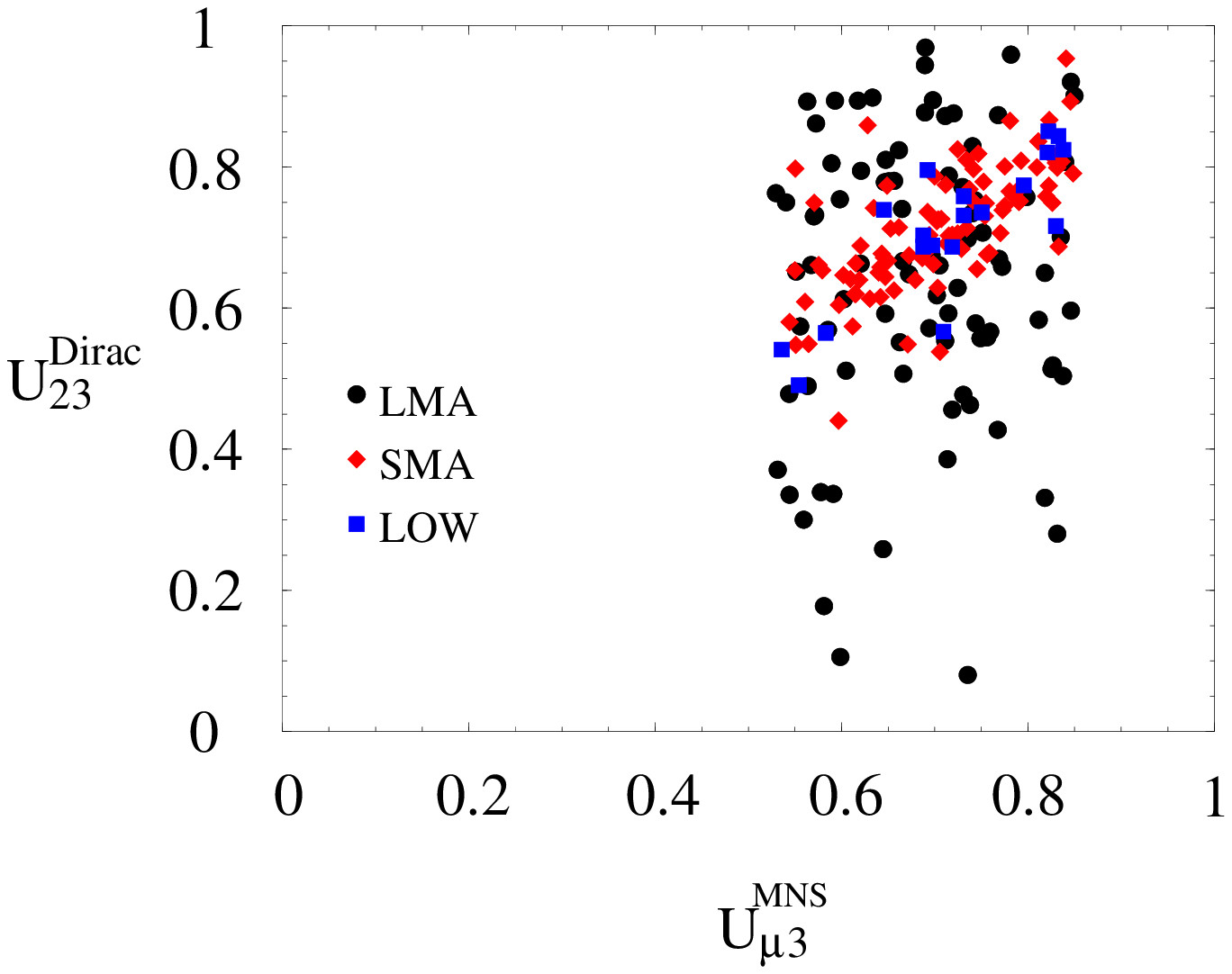,width=0.8\textwidth}
}
\centerline{
\psfig{figure=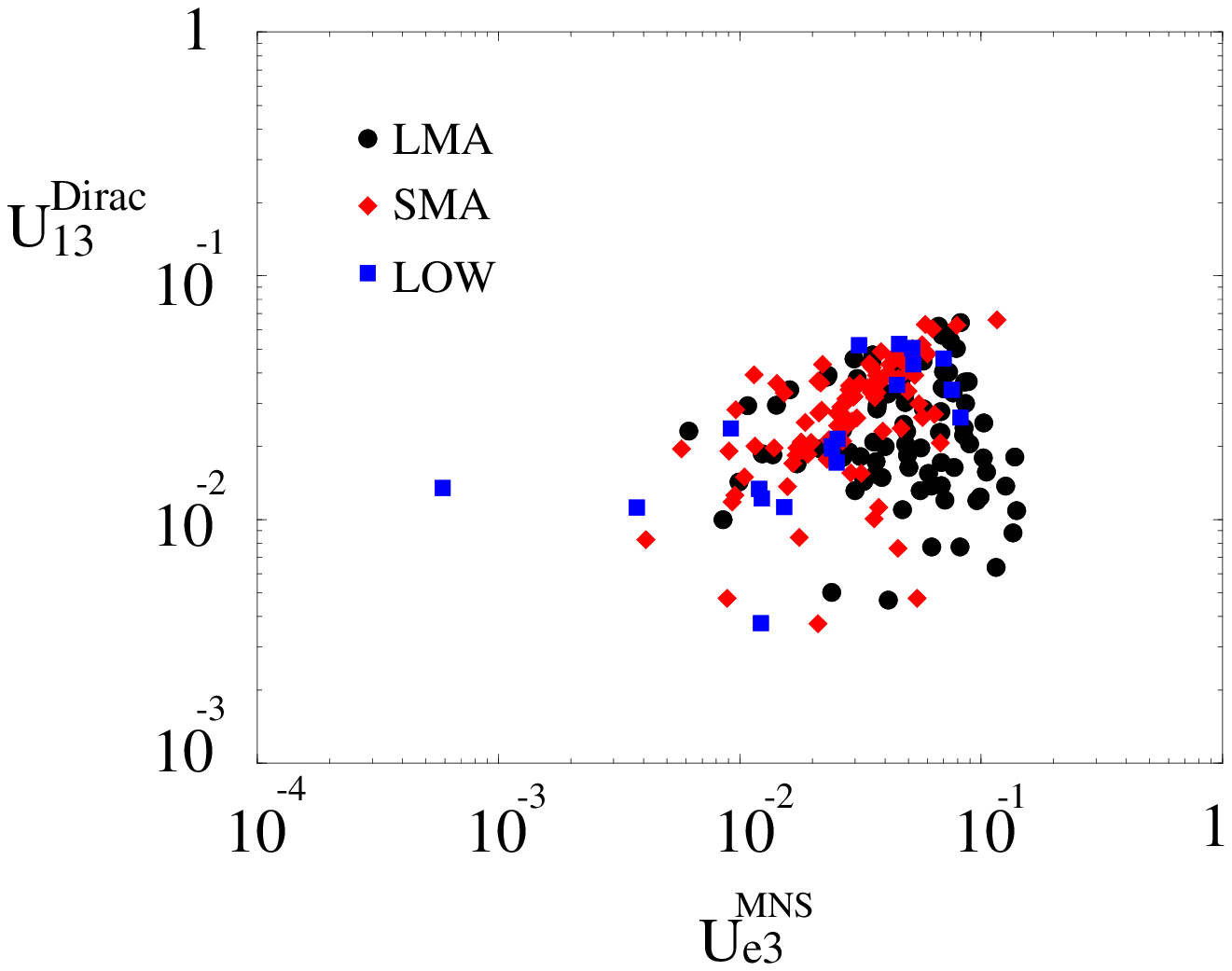,width=0.8\textwidth}
}
\caption{$U^{MNS}$ versus $U^{Dirac}$ in Model I 
with $(a,b,c)=(0,0,0)$.}
\label{Vd_Vn_13_000}
\end{figure}

In the present case, the feature of the LFV is very similar to the
previous case: Model I with $(a,b,c)=(0,1,2)$. 
The hierarchy of neutrino Yukawa couplings is
$f_{\nu1}:f_{\nu2}:f_{\nu3} \sim \epsilon:\epsilon-1:1$, which
depends on the solar neutrino solutions, and hence also the contributions
induced by $f_{\nu 2}$ will be important.
The leading contribution to the LFV mass $(\Delta m^2_{\tilde{L}})_{32}$
is given by
\begin{eqnarray}
(\Delta m^2_{\tilde{L}})_{32} &\simeq& 
-\frac{(6+a_0^2)m_0^2}{16 \pi^2} \left[
U^{Dirac}_{33} U^{Dirac*}_{23} |f_{\nu3}|^2 
\right.
\nonumber \\ 
&& \left. \hspace{1.5cm} +
U^{Dirac}_{32} U^{Dirac*}_{22} |f_{\nu2}|^2 \right]
\log \frac{M_G}{M_R},
\nonumber 
\\
&\sim& -\frac{(6+a_0^2)m_0^2}{16 \pi^2} 
 U^{Dirac}_{33} U^{Dirac*}_{23}\left(
|f_{\nu3}|^2 -|f_{\nu2}|^2 \right) 
\log \frac{M_G}{M_R}.
\label{ml_32}
\end{eqnarray}
In the SMA and LOW solutions,
$|U^{Dirac}_{33} U^{Dirac}_{23}|\sim 0.5$ because of the lopsided 
structure of the Dirac neutrino mass matrix, 
and 
$(|f_{\nu3}|^2-|f_{\nu2}|^2)/(|f_{\nu1}|^2+|f_{\nu2}|^2+|f_{\nu3}|^2)\sim1$
since the mass scale for solar neutrinos is much smaller than
the mass scale for atmospheric neutrinos 
(i.e. $|f_{\nu3}|\gg|f_{\nu2}|,~|f_{\nu1}|$). 
On the other hand, in the LMA solution,
$|U^{Dirac}_{33} U^{Dirac}_{23}|$ and 
$(|f_{\nu3}|^2-|f_{\nu2}|^2)/(|f_{\nu1}|^2+|f_{\nu2}|^2+|f_{\nu3}|^2)$ can
be widely spread since other matrix elements of $U^{Dirac}$
may have a large mixing and the mass scale for solar neutrinos can be close to
the scale for atmospheric neutrinos. 
In Fig.~\ref{Vd_Vn_13_000}, for example, we show the distribution of 
the values for $U^{Dirac}_{23}$ compared with the values for $U^{MNS}_{23}$.
In the SMA and LOW solutions, $U^{Dirac}_{23}\simeq U^{MNS}_{\mu 3}$; on the
other hand, the values for $U^{Dirac}_{23}$ in the LMA solution are widely 
distributed.
Therefore, the values of the branching 
ratio for $\tau \rightarrow \mu \gamma$ in the case of the 
LMA solution can be 
much more broadly distributed than in the other cases.

The dominant contribution to the LFV mass $(\Delta m^2_{\tilde{L}})_{21}$
can be written as
\begin{eqnarray}
(\Delta m^2_{\tilde{L}})_{21} &\simeq&
-\frac{(6+a_0^2)m_0^2}{16 \pi^2} \left[
U^{Dirac}_{23} U^{Dirac*}_{13} |f_{\nu3}|^2 
\right.
\nonumber \\ 
&& \left. \hspace{1.5cm} +
U^{Dirac}_{22} U^{Dirac*}_{12} |f_{\nu2}|^2 \right]
\log \frac{M_G}{M_R},
\nonumber \\
&=& -\frac{(6+a_0^2)m_0^2}{16 \pi^2} \left[
U^{Dirac}_{23} U^{Dirac*}_{13}\left( |f_{\nu3}|^2
-|f_{\nu2}|^2 \right)
\right. \nonumber \\
&&\left. \hspace{2.5cm} -U^{Dirac}_{21} U^{Dirac*}_{11}|f_{\nu2}|^2 \right]
\log \frac{M_G}{M_R}.
\label{ml_21}
\end{eqnarray}
In the cases of the SMA and LOW solutions, the first term 
in Eq.~(\ref{ml_21}) is
dominant since $|f_{\nu3}| \gg |f_{\nu2}|$ and $U^{Dirac}_{13}$
is non-negligible, as shown in Fig.~\ref{Vd_Vn_13_000}.
In the case of the LMA solution,
the second term can be large since $f_{\nu3} \sim f_{\nu2}$ in order to
get mass scale for the LMA solution, and $U^{Dirac}_{21}$ can be
large because of the large mixing for the LMA solution.
The matrix elements
$|U^{Dirac}_{23} U^{Dirac}_{13}|$ and $|U^{Dirac}_{21} U^{Dirac}_{11}|$
are broadly distributed (for example, see in Fig.~\ref{Vd_Vn_13_000}), 
and hence the value of the branching ratio for
$\mu \rightarrow e \gamma$ can be widely distributed in all cases.

When $|f_{\nu3}|^2 \gg |f_{\nu2}|^2$, a ratio of branching ratios can be
approximately as follows:
\begin{eqnarray}
\frac{{\rm Br}(\mu \rightarrow e \gamma)}
{{\rm Br}(\tau \rightarrow \mu \gamma)} &\sim&
\frac{1}{0.17} \left|\frac{(\Delta m^2_{\tilde{L}})_{21}}
{(\Delta m^2_{\tilde{L}})_{32}}\right|^2
\sim 0.01 \left|\frac{U^{Dirac}_{13}/0.03}{U^{Dirac}_{33}/0.7}\right|^2.
\label{ratio_branch_000}
\end{eqnarray}
Taking into account the present experimental limits 
and the future expectation on these processes, 
the search for $\mu \rightarrow e \gamma$
can be much more sensitive to this class of models than the search for 
$\tau \rightarrow \mu \gamma$.

In the following subsections, 
in order to obtain the magnitude of the branching ratios for the LFV
processes, we consider two cases again: Case 1, $\tau=0$ and
Case 2, $\tau=1$.

\subsubsection{Case 1, $\tau=0$}
\begin{figure}
\centerline{
\psfig{figure=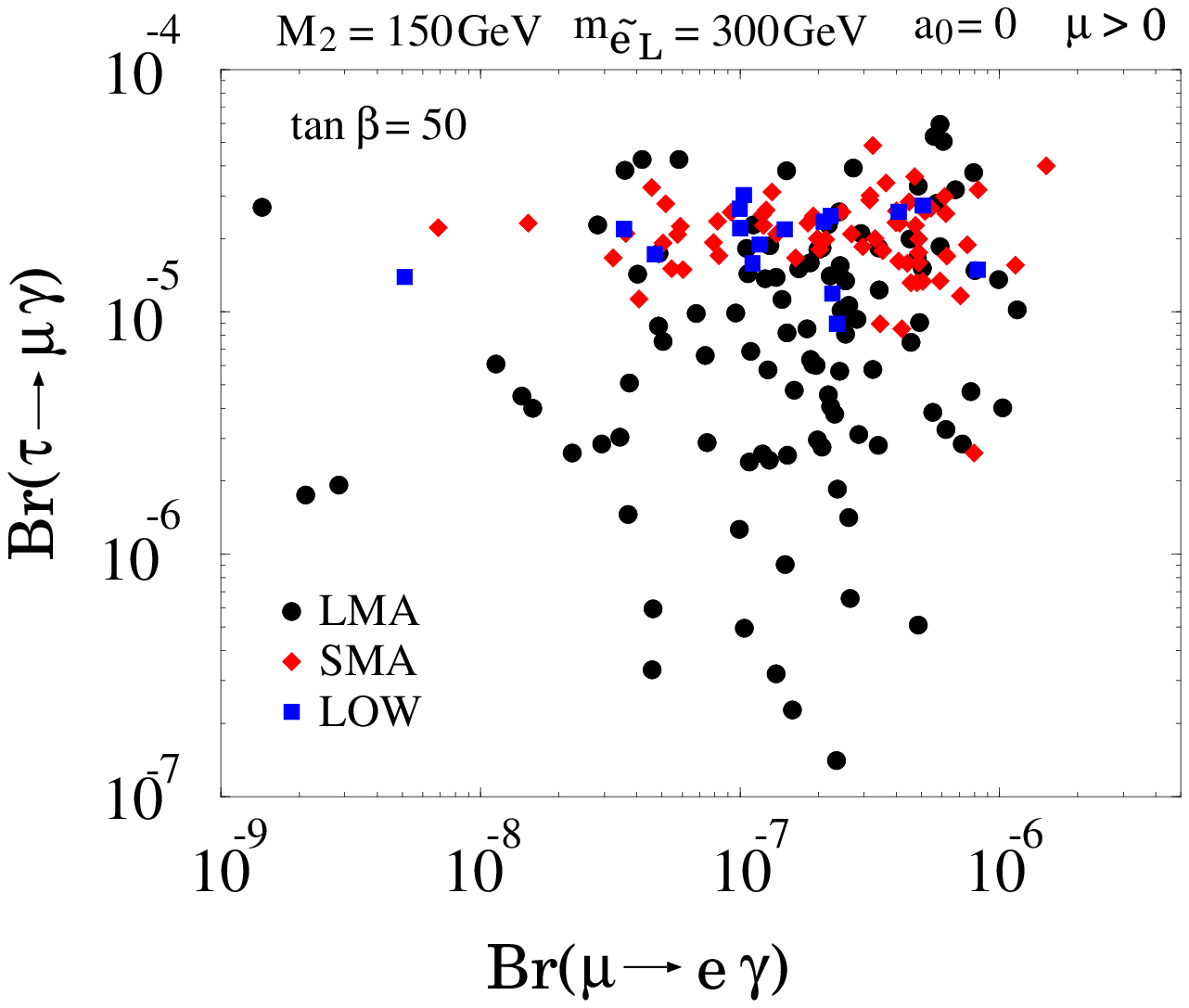,width=0.73\textwidth}
}
\caption{Br($\mu \rightarrow e \gamma$) versus
Br($\tau \rightarrow \mu \gamma$) in Model I with $(a,b,c,\tau)=(0,0,0,0)$.
Here we take the left-handed slepton mass to be 300 GeV, the Wino mass
to be 150 GeV, and $\epsilon=0.07$.}
\label{lfv_case000_tan50}
\vspace*{0.2cm}
\centerline{~~~
\psfig{figure=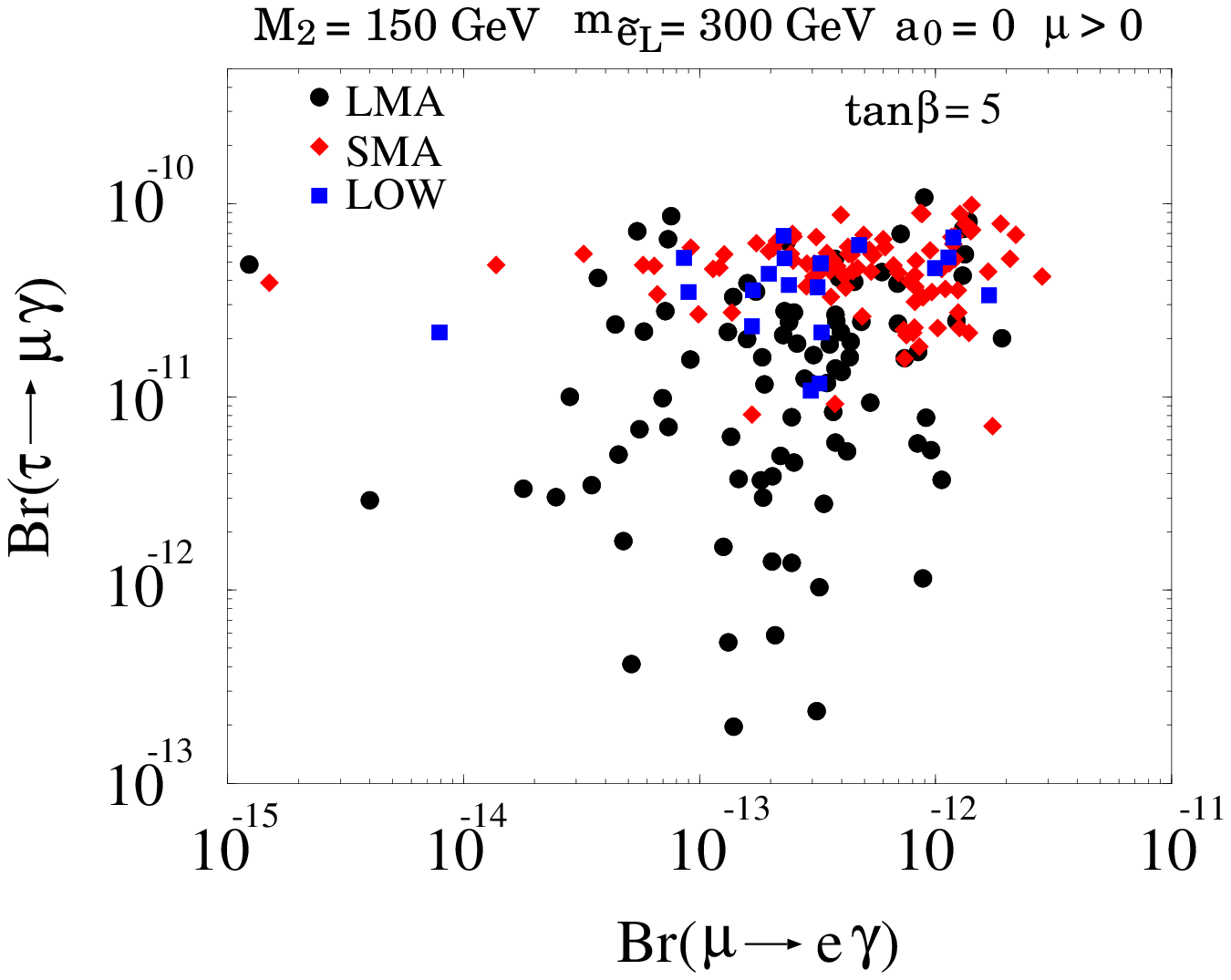,width=0.78\textwidth}
}
\vspace*{-0.3cm}
\caption{Same as Fig.\ref{lfv_case000_tan50} except for
Model I with $(a,b,c,\tau)=(0,0,0,1)$.}
\label{lfv_case000_tan5}
\end{figure}

As in the previous models in section 4.1.1,
we take $\tan \beta=50$ in order for all the third-generation
Yukawa couplings to be of order 1. We assume that $m_3=m_{top}$ at the
GUT scale
in Eq.~(\ref{dirac_mass_model1_000}) with $\tau=0$.
In Fig.~\ref{lfv_case000_tan50} we show the result of the branching
ratios for the $\tau \rightarrow \mu \gamma$ and $\mu \rightarrow e \gamma$
processes. 
Because of the large neutrino Yukawa coupling, the branching ratios of the 
$\mu \rightarrow e \gamma$ process are so large that most of the parameter 
region can be excluded by the present limit on $\mu \rightarrow e \gamma$.
As in the models with the FN charges $(a,b,c)=(0,1,2)$, the models with 
$\tau=0$, which have a mild Yukawa unification between the top and
tau-neutrino, are almost excluded by the bound on $\mu \rightarrow e \gamma$
\cite{JTY}.

\subsubsection{Case 2, $\tau=1$}
In the models with $\tau =1$, as discussed in section 4.1.2, 
the following relations among the third generation
Yukawa couplings are implied: $f_b\sim f_\tau \sim f_{\nu 3}
\sim \epsilon f_{top}$. Thus we assume that in 
Eq.~(\ref{dirac_mass_model1_000}) with $\tau=1$, $m_3=m_{top}$ at 
the GUT scale and
take $\tan\beta$ to be 5 in order to realize the relation.
As compared to the case with $\tau=0$, the branching ratios for
$\tau \rightarrow \mu \gamma$ and $\mu \rightarrow e \gamma$ are
suppressed by $\epsilon^4$, with the additional suppression due to small
$\tan\beta$.
In Fig.~\ref{lfv_case000_tan5} we show the result of the branching ratios
for $\tau\rightarrow \mu \gamma$ and $\mu \rightarrow e \gamma$.
For the $\mu \rightarrow 3e$ and $\mu \rightarrow e$ conversion processes,
we have checked that the relations in Eqs.~(\ref{mu_3e}) and
(\ref{mu_e_conv}) are
held. Therefore, interestingly, the predicted event rates
for $\mu \rightarrow e \gamma$ and $\mu \rightarrow e$ conversion can be
as large as those the future experiments can reach.
On the other hand, for $\tau \rightarrow \mu \gamma$ process,
the significant improvement of the branching
ratio sensitivity of much below $10^{-10}$
will be needed in order to reach the predicted values.

\subsection{Model II ($\delta=0$) with $(a,b,c,\tau)=(0,1,2,1)$}
\begin{figure}
\vspace*{-0.5cm}
\centerline{
\psfig{figure=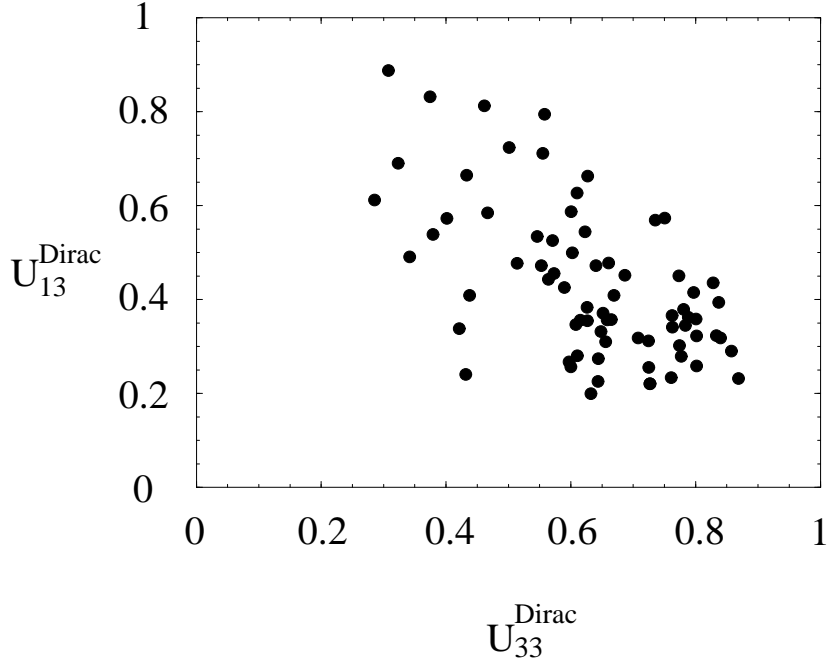,width=0.8\textwidth}
}
\caption{$U^{Dirac}_{13}$ versus $U^{Dirac}_{33}$ in Model II
with $(a,b,c)=(0,1,2)$.}
\label{V13_33_model2}
\end{figure}

Let us discuss the LFV in Model II ($\delta=0$).
First we consider the case with hierarchical right-handed neutrino masses,
that is the case with $(a,b,c)=(0,1,2)$.
The Dirac neutrino mass matrix is given by
\begin{eqnarray}
m_{\nu D} = m_3 \epsilon \left(
\begin{array}{ccc}
\bar{C}_3 \epsilon^2 & \bar{B}_3 \epsilon^2 & \bar{A}_3 \epsilon^2 \\
\bar{C}_2 \epsilon & \bar{B}_2 \epsilon  & \bar{A}_2 \epsilon \\
\bar{C}_1    & \bar{B}_1  & \bar{A}_1 
\end{array}
\right).
\end{eqnarray}
Since the matrix elements $(m_{\nu D})_{3 i}/(m_3 \epsilon)$ ($i=1$--3)
are of order 1, we can expect that all elements of the mixing matrix
$U^{Dirac}$ would be of order one: $U^{Dirac}_{ij} \sim O(1)$, even
if we imposed the constraints of neutrino parameters on the neutrino
mixing matrix $U^{MNS}$.
This structure of the mixing matrix $U^{Dirac}$ is quite distinct
from that in Model I, as seen in Eqs.~(\ref{dirac_mix_matrix}) and
(\ref{Vdirac_000}).
For example. in Fig.~\ref{V13_33_model2}, 
we show $U^{Dirac}_{13}$ compared with $U^{Dirac}_{33}$.
Even though the CHOOZ limit $|U^{MNS}_{e3}|<0.15$ was imposed, 
the component $U^{Dirac}_{13}$
can be of order 1. This large $U^{Dirac}_{13}$ is a very important
feature for the LFV in the present models.

Because of the FN charges of right-handed neutrinos, the structure of 
neutrino Yukawa couplings is hierarchical ($f_{\nu1}:f_{\nu2}:f_{\nu3}
\sim \epsilon^2 :\epsilon :1$). Therefore, only the third-generation
Yukawa coupling $f_{\nu 3}$ is important for contributions 
to LFV masses. The dominant terms
of the matrix elements ($(\Delta m^2_{\tilde{L}})_{32}$)
and ($(\Delta m^2_{\tilde{L}})_{21}$) are given by
\begin{eqnarray}
(\Delta m^2_{\tilde{L}})_{32} &\simeq& -\frac{(6+a_0^2)m^2_0}
{16 \pi^2} U_{33}^{Dirac} U_{23}^{Dirac *} |f_{\nu 3}|^2 
\log \frac{M_G}{M_R},
\\
(\Delta m^2_{\tilde{L}})_{21} &\simeq& -\frac{(6+a_0^2)m^2_0}
{16 \pi^2} U_{23}^{Dirac} U_{13}^{Dirac *} |f_{\nu 3}|^2 
\log \frac{M_G}{M_R}.
\label{21_comp_model2}
\end{eqnarray}
Then a ratio of branching ratios 
Br($\mu\rightarrow e \gamma$)/Br($\tau \rightarrow \mu \gamma$)
are approximately expressed as follows:
\begin{eqnarray}
\frac{{\rm Br}(\mu \rightarrow e \gamma)}
{{\rm Br}(\tau\rightarrow \mu \gamma)} \simeq
\frac{1}{0.17}\left|
\frac{(\Delta m^2_{\tilde{L}})_{21}}
{(\Delta m^2_{\tilde{L}})_{32}} \right|^2 \sim
6 \left|\frac{U^{Dirac}_{13}}{U^{Dirac}_{33}} \right|^2.
\end{eqnarray}
Since $|U^{Dirac}_{13}|\sim |U^{Dirac}_{33}|$, as shown in 
Fig.~\ref{V13_33_model2}, the branching ratio
for $\mu \rightarrow e \gamma$ can be as large as
or even larger than that for $\tau \rightarrow \mu \gamma$.
As discussed in Model I, the models with $\tau=0$, in which
a mild Yukawa unification between the top and tau-neutrino
is realized, are almost excluded by the present $\mu \rightarrow e \gamma$
bound~\cite{JTY}. Therefore, in this section, we only present the result
in the models with $\tau=1$.

\begin{figure}[t]
\centerline{
\psfig{figure=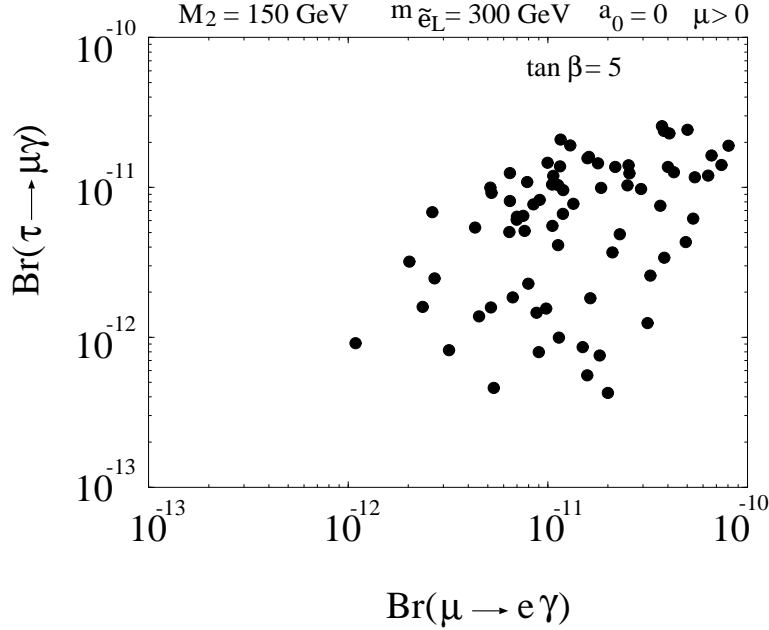,width=0.75\textwidth}
}
\caption{Br($\mu \rightarrow e \gamma$) versus
Br($\tau \rightarrow \mu \gamma$) in Model II with $(a,b,c,\tau)=(0,1,2,1)$.
Here we take the left-handed slepton mass to be 300 GeV, the Wino mass
to be 150 GeV, and $\epsilon=0.07$.}
\label{lfv_case_mode2_tan5}
\end{figure}

In Fig.~\ref{lfv_case_mode2_tan5}, we show the numerical result
for these branching ratios.
The predicted branching ratios for $\mu \rightarrow e \gamma$
are larger than those in Model I, since the mixing element
$U^{Dirac}_{13}$ is much larger than that in the Model I.
On the other hand, the branching ratios for $\tau \rightarrow \mu \gamma$
are almost the same as those in Model I.
\footnote{If the inverted neutrino mass hierarchy is realized, as discussed 
in section 3.2, neutrino Yukawa couplings $f_{\nu 1}$ and $f_{\nu 2}$
can be so large that they can contribute to the LFV masses. In this case,
we have checked that the branching ratios for the LFV processes, especially 
for $\mu \rightarrow e \gamma$, could be more enhanced
because of the additional contributions from  $f_{\nu 1}$ and $f_{\nu 2}$
to the LFV masses.}
As expected, the predicted branching ratios for $\mu \rightarrow e \gamma$
are as large as those for $\tau \rightarrow \mu \gamma$.
Note that for the $\mu \rightarrow 3e$ and $\mu \rightarrow e$ conversion
processes, the relations in Eqs.~(\ref{mu_3e}) and (\ref{mu_e_conv}) are
predicted.
Therefore  the search for LFV in muon processes is much more
sensitive to this class of models too.
Even at present, some of the points in Fig.~\ref{lfv_case_mode2_tan5}
are already excluded by the current
experimental limit on $\mu \rightarrow e \gamma$ 
[Br$(\mu \rightarrow e \gamma)<1.2 \times 10^{-11}$~\cite{PDG}].
The future proposed improvement of the limit on LFV in muon processes 
will definitely provide a significant test of this class of models.

\subsection{Model II ($\delta=0$) with $(a,b,c,\tau)=(0,0,0,1)$}
Next we consider the models in which the heavy right-handed neutrinos
are nearly degenerate, that is the models with $(a,b,c)=(0,0,0)$.
Again here we only discuss the case with $\tau =1$.

The Dirac neutrino mass
matrix is given by
\begin{eqnarray}
m_{\nu D} = m_3 \epsilon \left(
\begin{array}{ccc}
\bar{C}_3  & \bar{B}_3  & \bar{A}_3  \\
\bar{C}_2  & \bar{B}_2  & \bar{A}_2  \\
\bar{C}_1  & \bar{B}_1  & \bar{A}_1 
\end{array}
\right).
\label{dirac_model2_000}
\end{eqnarray}
\begin{figure}
\vspace*{-0.5cm}
\centerline{
\psfig{figure=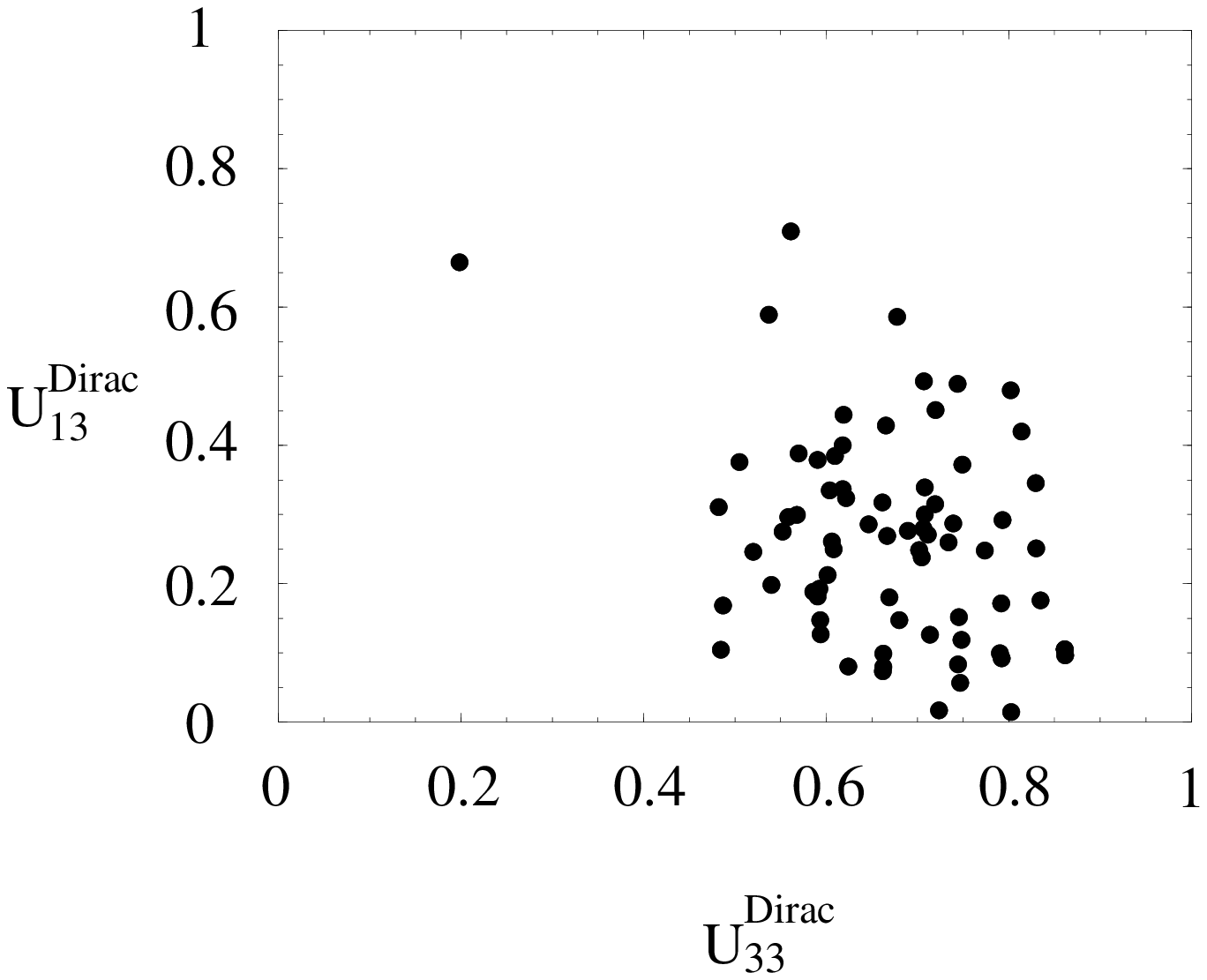,width=0.75\textwidth}
}
\caption{$U^{Dirac}_{13}$ versus $U^{Dirac}_{33}$ in Model II
with $(a,b,c)=(0,0,0)$.}
\label{V13_33_model2_000}
\end{figure}
\begin{figure}[t]
\centerline{
\psfig{figure=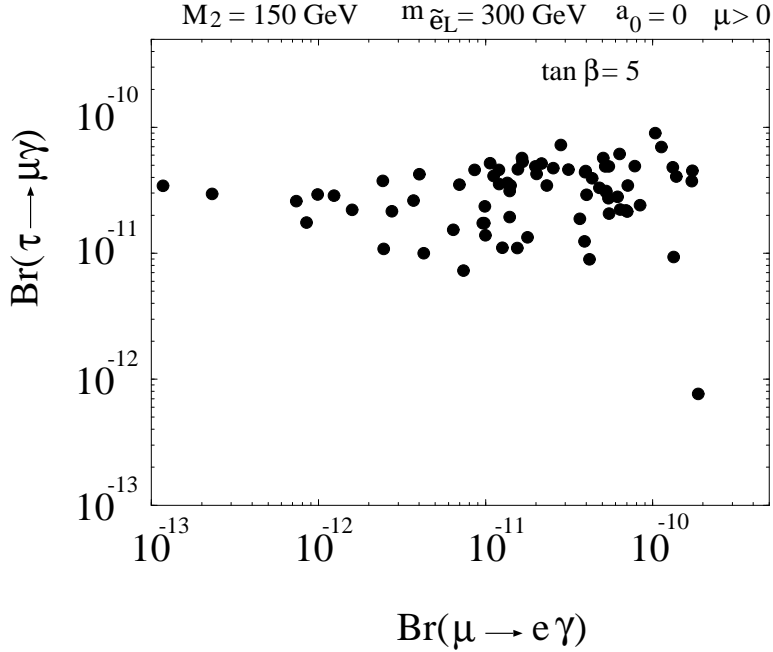,width=0.75\textwidth}
}
\vspace*{-0.3cm}
\caption{Br($\mu \rightarrow e \gamma$) versus
Br($\tau \rightarrow \mu \gamma$) in Model II with $(a,b,c,\tau)=(0,0,0,1)$.
Here we take the left-handed slepton mass to be 300 GeV, the Wino mass
to be 150 GeV, and $\epsilon=0.07$.}
\label{lfv_case_model2_000_tan5}
\end{figure}
As in the previous case of Model II, we expect 
$U^{Dirac}_{ij}\sim O(1)$.
In this case, if coefficients $\bar{A}_i$, $\bar{B}_i$ and $\bar{C}_i$
are all real, $U^{Dirac}=U^{MNS}$. Thus, as shown in 
Fig.~\ref{V13_33_model2_000}, 
the component $U^{Dirac}_{13}$
tends to be smaller than that in the previous case,
because the constraint $|U^{MNS}_{e3}|<0.15$ slightly affects
the value of $U^{Dirac}_{13}$.
However, all components in the Dirac mass
matrix Eq.~(\ref{dirac_model2_000}) are of order one, and all Yukawa
couplings $f_{\nu i}~(i=1$--3) can contribute to the LFV in slepton masses.
Therefore, branching ratios of  $\tau \rightarrow \mu \gamma$ and
$\mu \rightarrow e \gamma$ can be larger than those in the previous case
of Model II.
We present the numerical result in Fig.~\ref{lfv_case_model2_000_tan5}.
As can be seen in Fig.~\ref{lfv_case_model2_000_tan5}, 
the branching ratio for $\mu \rightarrow e \gamma$
can be as large as that of $\tau \rightarrow \mu \gamma$, because
all components $V^{Dirac}$ can be of the same order. Many of the points
are already excluded by the present experimental limit on
the $\mu \rightarrow e \gamma$ process, and the future improvement will
be able to test this class of models.

\section{Conclusions}
We discussed the neutrino masses and LFV in SUSY models with
lopsided FN  charges. For neutrino physics, we stressed the importance
of the measurement of $U^{MNS}_{e3}$ in both Models I and II.
In Model II, the LMA solution is strongly preferred so that
the near-future KamLAND experiment could play a significant role
for this class of models.

We also analyzed the LFV in detail within this framework.
The present experimental limits on LFV processes almost
excluded the models with $\tau=0$ in which a mild Yukawa coupling
unification between the top and tau-neutrino is realized at the GUT scale.
In the models with $\tau=1$, in which the tau-neutrino Yukawa coupling
is suppressed by $\epsilon$ compared with the top Yukawa coupling,
the predicted branching ratios for $\tau \rightarrow \mu \gamma$ 
are as large as or much less than $10^{-10}$, and hence a significant
experimental improvement of the limit on 
the branching ratio is needed in order to
reach the predictions of this framework.
On the other hand, the predicted branching ratios
for $\mu \rightarrow e \gamma$ and $\mu \rightarrow e$ conversion in nuclei
can be as large as those the future proposed experiments can reach.
Therefore, the future experiments can provide the significant
test of this realistic framework for neutrino masses 
as well as charged lepton and quark masses.
\section*{Acknowledgments}
We are greatly indebted to Tsutomu Yanagida for his collaboration at an 
early stage. The work of J.S. is supported  in part by a Grant-in-Aid 
for Scientific Research of the Ministry of Education, Science and Culture,
\#12047221, \#12740157.
%
%
%
%
%
\newcommand{\Journal}[4]{{#1} {\bf #2}, {#4} {(#3)}}
\newcommand{\APJ}{Ap. J.}
\newcommand{\CJP}{Can. J. Phys.}
\newcommand{\EPJ}{Eur. Phys. J.}
\newcommand{\MPL}{Mod. Phys. Lett.}
\newcommand{\NC}{Nuovo Cimento}
\newcommand{\NP}{Nucl. Phys.}
\newcommand{\PL}{Phys. Lett.}
\newcommand{\PR}{Phys. Rev.}
\newcommand{\PRep}{Phys. Rep.}
\newcommand{\PRL}{Phys. Rev. Lett.}
\newcommand{\PTP}{Prog. Theor. Phys.}
\newcommand{\SJNP}{Sov. J. Nucl. Phys.}
\newcommand{\ZP}{Z. Phys.}
\newcommand{\EUR}{Eur. Phys. J.}

\end{document}